\documentclass[preprint,12pt]{elsarticle}
\usepackage{lineno,hyperref}
\modulolinenumbers[5]
\usepackage{graphicx}
\usepackage{amssymb}
\usepackage{epstopdf}
\usepackage{amsmath}
\usepackage{bm}
\usepackage{float}
\journal{Comput. Phys. Comm. (BU-HEPP-19-02)}
\begin{document}

\title{Disconnected Loop Subtraction Methods in Lattice QCD}


\author[Bay1,Everst]{Suman Baral}
\ead{suman\_baral@baylor.edu, suman@neuralinnovations.io}
\author[Bay1]{Travis Whyte}
\ead{travis\_whyte@baylor.edu}
\author[Bay1]{Walter Wilcox}
\ead{walter\_wilcox@baylor.edu}
\author[Bay2]{Ronald B. Morgan}
\ead{ronald\_morgan@baylor.edu}
\address[Bay1]{Department of Physics, Baylor University, Waco, TX 76798-7316}
\address[Bay2]{Department of Mathematics, Baylor University, Waco, TX 76798-7316}
\address[Everst]{Everest Institute of Science and Technology, Samakhusi Kathmandu, Nepal}









\begin{abstract}
Lattice QCD calculations of disconnected quark loop operators are extremely computer time-consuming to evaluate. To compute these diagrams using lattice techniques, one generally uses stochastic noise methods. These employ a randomly generated set of noise vectors to project out physical signals. In order to strengthen the signal in these  calculations, various noise subtraction techniques may be employed. In addition to the standard method of perturbative subtraction, one may also employ matrix deflation techniques using the GMRES-DR and MINRES-DR algorithms as well as polynomial subtraction techniques to reduce statistical uncertainty. Our matrix deflation methods play two roles: they both speed up the solution of the linear equations as well as decrease numerical noise. We show how to combine deflation with either perturbative and polynomial methods to produce extremely powerful noise suppression algorithms. We use a variety of lattices to study the effects. In order to set a benchmark, we first use the Wilson matrix in the quenched approximation. We see strong low eigenmode dominance at kappa critical ($\kappa_{crit}$) in the variance of the vector and scalar operators. We also use MILC dynamic lattices, where we observe deflation subtraction results consistent with the effectiveness seen in the quenched data.
\end{abstract}


                              

\maketitle

\section{Introduction}

\begin{figure}[h]\label{Fone}
\centering
\includegraphics[trim={0 0cm 0 0cm},clip,width=.90\textwidth]{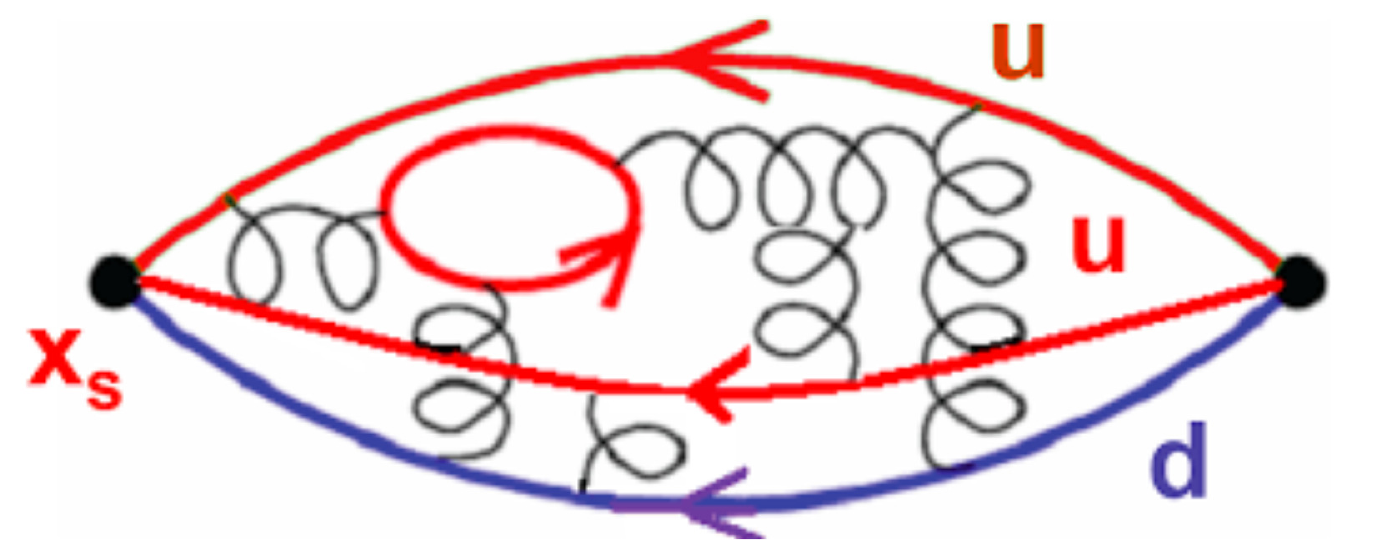}
\caption{An illustrative disconnected quark loop within a proton. The solid lines represent various flavors of quarks, the loopy lines represent the gluons. The quark loop is connected to the rest of the diagram only by gluon lines. The final position $x_s$ may be summed over to produce a proton at rest.}\label{figg1}
\end{figure}

Lattice QCD (LQCD) is a set of numerical techniques which use a finite space-time lattice to simulate the interactions between quarks and gluons. Such amplitudes are affected by the background of quark-antiquark loops in particles such as a proton. The evaluation of loop effects on a given lattice, such as is illustrated in Fig.~\ref{figg1}, is extremely computer time intensive and approximation techniques must be introduced. Such contributions are termed \lq\lq disconnected diagrams", but more correctly these loops are connected to the other parts of the amplitude by additional gluon lines, and their effects are contained in loop/amplitude configuration correlations. Quark loops can form a scalar, pseudoscalar, vector, pseudovector or tensor operator. In this work we concentrate on lattice operators which respond well to unpartitioned noise\cite{Wilcox:1999ab}: scalar and vector operators. Most physical amplitudes in QCD are affected by such loop effects; some examples include nucleon electromagnetic form factors~\cite{Alexandrou:2012zz,Alexandrou:2018sjm}, the strangeness and charm contents of the nucleon~\cite{wilcoxtm,Sufian:2016pex,strange}, the determination of the mass of flavor singlet mesons\cite{Jansen:2008wv}, multiquarks and scattering states\cite{Bulava:2008qx}, hadronic scattering lengths and structure functions\cite{scatter,Deka:2008xr}, and electron or muon hadronic $g-2$ loop contributions\cite{Lepage,Feng:2011zk}. To tackle such difficult lattice problems one projects out operator expectation values using unbiased noise stochastic estimates\cite{Bernardson1993he,noisestat1,noisestat2}. Subtraction methods are then applied to reduce the statistical uncertainty of these noisy operators\cite{Wilcoxpert,vicpaper}. Perturbative subtraction\cite{Bernardson1993he,pert1} is a standard noise subtraction method which is often used in this context.

We are continuing to build upon our previous work in the area of improving quark loop calculations. In these works we have shown the effectiveness of the linear equations/deflation algorithms Generalized Minimum Residual-Deflated Restarted (GMRES-DR) and Minimum Residual-Deflated Restarted (MINRES-DR) in an LQCD context\cite{GMRES-DR,MINRESDR}. In addition, we have introduced noise suppression techniques called polynomial matrix methods\cite{polynomial,Liu2} and eigenspectrum deflation methods\cite{Morgan:2001qr,Darnell:2003mi,Wilcox2,Nicely,Baral}. In \cite{vicpaper,Guerrero2} we also introduced techniques which combine deflation with other subtraction methods. The overall goal is to improve the efficiency of LQCD noise method algorithms, especially with respect to perturbative subtraction at low quark mass. In this regard one would expect this standard method, which involves a simple perturbative expansion in the quark hopping parameter of the quark matrix, to lose its effectiveness as the quark mass is decreased. Indeed, this is what we observe in our simulations. In sharp contrast, however, we also see a dramatic increase in the effectiveness of deflation methods in the same limit, and combination techniques, integrating perturbative or polynomial expansions with hermitian deflation, are even more effective. Our matrix deflation methods play a double role: they both speed up the solution of the linear equations as well as decrease the numerical noise. We will see that both our iterative (in the form of MINRES-DR) and subtractive (in the form of Hermitian Forced Eigenvalue Subtraction (HFES), Hermitian Forced Perturbative Subtraction (HFPS), and Hermitian Forced Polynomial Subtraction (HFPOLY)) methods work effectively all the way down to the zero quark mass limit. Our emphasis here is on improving simulation techniques for LQCD at low quark mass, and for this purpose we find it convenient to primarily consider QCD matrices in the so-called quenched approximation, where quark loops are not included in the background configurations. However, we will also consider a set of dynamical lattices from MILC~\cite{MILC}, where we will observe results consistent with the effectiveness seen in the quenched data.

There have been many other approaches to reducing variance in LQCD disconnected diagram loop calculations. One of these techniques is called the truncated solver method\cite{deForcrand:1998je,Bali:2009hu} and consists of stopping the inversion of quark propagators early in an unbiased way. Related to this is the method referred to as truncated eigenmode acceleration\cite{Neff:2001zr,DeGrand:2002gm,Bali:2005pb}, where low eigenvalue information of the quark matrix is calculated separately and supplemented with stochastic estimates in the orthogonal space. Another of these approaches was referred to above and is called noise partitioning or dilution\cite{Wilcox:1999ab,Babich,Morningstar}. Even/odd checkerboarding is one such strategy, and spin/Dirac space partitioning is also an effective strategy for lattice operators such as quark pseudoscalar or pseudovector combinations. Another related approach is called hierarchical probing (HP)\cite{Green,Stathopoulos}. In this method the Euclidean noise vectors for the lattice are chosen to zero increasingly large distances on a checkboarded lattice. In Ref.~\cite{Gambhir}, a combination of HP and deflation is considered and a synergy of the combined methods is observed. We will use completely different methods here which do not depend on dilution. In contrast to Ref.~\cite{Gambhir} which uses up to 512 dilution vectors for each complete inversion solve, we use an adjustable number of undiluted noise vectors on a given configuration and require only 1 or 2 special super-convergence right-hand side solves to produce any desired matrix element.

We will start with a brief review of noise theory in Sections~\ref{Stwo} and \ref{Sthree}. The basic strategy of noise subtraction techniques is introduced in Section~\ref{Sfour} where we will see that off-diagonal components of matrices are responsible for the noise variance. The various methods we will consider will be discussed individually in Section~\ref{Sfive} and the lattice loop operators will be examined in Section \ref{Ssix}. In Section \ref{Sseven} we present the numerical test results for the various subtraction methods in both graphical and tabular forms. We also investigate two separate important cases: in Section~\ref{Seight} we consider the effect of subtraction {\em at} $\kappa_{critical}$ in the Wilson matrix case on large lattices and in Section~\ref{Snine} we examine subtraction techniques in the context of dynamical lattices obtained from the MILC collaboration. A summary and conclusion is made in Section~\ref{Sten} based on our numerical tests.

\section{Noise Methods}\label{Stwo}
Information about a system can be extracted by projecting its signal on the randomly generated noise vectors. For example, a QCD matrix $M$ can be projected over a randomly chosen noise vector $\eta^{(n)}$ to extract the solution vector $x$ as we can see in the equation,
\begin{equation}\label{xyz1}
\begin{aligned}
Mx^{(n)} = \eta^{(n)},
 \end{aligned}
\end{equation}
where $M$ is of size $\mathcal{N}\times\mathcal{N}$, $x^{(n)}$ and $\eta^{(n)}$ are of size $\mathcal{N}\times 1$ in space-time-color-Dirac space and of size $N$ in noise space. If $\alpha_i^{(n)}$ and $\beta_i^{(n)}$ are two arbitrary vectors, then the $ {(ij)}^{th}$ element of the product  $\langle\alpha_i \beta_i\rangle$ is defined as
  \begin{equation}\label{xyz4}
 \begin{aligned}
 \langle \alpha_i\beta_j \rangle \equiv  \lim_{N \rightarrow \infty } \frac{1}{N} \sum_n^N \alpha_i^{(n)} \beta_j^{*(n)}.
 \end{aligned}
\end{equation}
Of course in a computational context $N$ is finite and the solutions we obtain are approximate.
For a sufficiently large value of ${N}$, if we apply Eq.~(\ref{xyz4})  to noise vectors, we obtain
\begin{equation}
 \langle\eta_i\eta_j\rangle= \lim_{N \rightarrow \infty } \frac{1}{N} \sum_n^N \eta_i^{(n)} \eta_j^{*(n)}=\delta_{ij}.\label{ww1}
\end{equation}  
In addition, the expectation value of noise vectors yields
\begin{equation}
\langle\eta_i\rangle= \lim_{N \rightarrow \infty } \frac{1}{N} \sum_n^N \eta_i^{(n)} =0.
\end{equation}
With these two properties of  noise, along with the indexed form of Eqs.~(\ref{xyz1}) and (\ref{xyz4}), we are able to determine each element from the quark matrix inverse, $M^{-1}$ as follows.   
\begin{equation}\label{xyz168}
\begin{aligned}
M_{ik}^{-1}&
 = \sum_jM_{ij}^{-1} \delta_{jk} \\& 
= \sum_jM_{ij}^{-1} \langle \eta_j \eta_k \rangle \\& 
 = \sum_jM_{ij}^{-1} \left(\lim_{N \rightarrow \infty }\frac{1}{N}{\sum_n^N}  {\eta_j^{(n)}} {\eta_k^{*(n)}} \right)\\&
= \lim_{N \rightarrow \infty }\frac{1}{N}{\sum_n^N} \left( \sum_j M_{ij}^{-1} {\eta_j^{(n)}}\right){\eta_k^{*(n)}} \\&
 =\lim_{N \rightarrow \infty } \frac{1}{N}{\sum_n^N} {x_i^{(n)}} {\eta_k^{*(n)}} \\&
=\langle x_i \eta_k\rangle .
 \end{aligned}
\end{equation}
For finite $N$ computer simulations are not exact and hence there is an associated noise error. We apply several techniques which attempt to minimize the statistical uncertainty which will be discussed further below. We have used $Z(N)$ noises, with $N=4$, for our computer simulations. 


 \section{Variance}\label{Sthree}

Many operators in LQCD simulations are flooded with noise. Several strategies could be applied to reduce the variance of these operators, which originates in the off-diagonal components of the associated quark matrix. One basic strategy is to mimic the off-diagonal elements of the inverse of the quark matrix with another traceless matrix, thereby maintaining the trace but with reduced statistical uncertainty. In this section, we will explain the statistical theory.

We define $X$ as the matrix made from projection of two noise vectors:
\begin{equation}
X_{ij}  \equiv  \frac{1}{N} \sum_n^N \eta_i^{(n)} \eta_j^{*(n)}.
\end{equation} 
%
Of course in the $N\rightarrow\infty$ limit, Eq.~(\ref{ww1}) holds and we have
\begin{equation}
\lim_{N \rightarrow \infty }Tr\left(QX\right)=Tr\left(Q\right).
\end{equation} 
As far as the variance of this quantity is concerned, it can be shown that\cite{Bernardson1993he,Wilcox:1999ab}
 \begin{equation}\label{xyz207}
 \begin{aligned}
V\left[Tr\left(QX\right)\right]& \equiv \langle | {\sum_{i}^N} q_{ij} X_{ji}- Tr(Q) |^2 \rangle \\&
= \sum_{i\not=j} \left(\langle |X_{ji}|^2\rangle |q_{ij}|^2 + q_{ij} q^*_{ji} \langle(X_{ji})^2\rangle\right)  +\sum_i \langle|X_{ii} - 1|^2 \rangle |q_{ii}|^2,
\end{aligned}
\end{equation}    
where $Q$ is the matrix-representation of an operator. 

First, let's consider a general real noise. The constraints are:
\begin{equation}\label{xyz230}
\begin{aligned}
\langle|X_{ji}|^2\rangle = 
\langle(X_{ji})^2\rangle = \frac{1}{N},
\end{aligned}
\end{equation}
for $i\neq j$. For general real noises variance can now be written as
\begin{equation}
\begin{aligned}
V\left[Tr(QX_{real})\right] =\frac{1}{N} \sum_{i\not=j} \left(|q_{ji}|^2 + q_{ij} q^*_{ji}\right) +\sum_i\langle|X_{ii}-1|^2\rangle |q_{ii}|^2.
\end{aligned}
\end{equation}
For $Z(2)$ noises apart from the constraints at Eq.~(\ref{xyz230}), there is one additional constraint, $\langle|X_{ii} - 1|^2 \rangle=0$ for $i=j$. This gives
\begin{equation}
\begin{aligned}
V\left[Tr\left(QX_{Z(2)}\right)\right]=\frac{1}{N} \sum_{i\not=j} \left(|q_{ji}|^2 + q_{ij} q^*_{ji}\right).
\end{aligned}
\end{equation}
For $Z(4)$ noises, the constraints are little different. One has the conditions
\begin{equation}
\begin{aligned}
& \langle|X_{ji}|^2\rangle  = \frac{1}{N}, \\&
\langle(X_{ji})^2\rangle =0, \\&
\langle|X_{ii}-1|^2\rangle =0,
\end{aligned}
\end{equation}
for $Z(N),N\ge 3$, which makes the variance for Eq.~(\ref{xyz207}) to be 
 \begin{equation}\label{xyz255}
\begin{aligned}
V\left[Tr\left(QX_{Z(N\ge3)}\right)\right] = \frac{1}{N} \sum_{i\not=j} |q_{ji}|^2 .
\end{aligned}
\end{equation}
From this equation we can see that the variance only depends on off-diagonal elements of the matrix for $Z(4)$ noises. This understanding is the foundation for the subtraction noise techniques that we develop.

\section{Subtraction Noise Techniques}\label{Sfour}

Noise subtraction technique can be used to reduce the error associated with  disconnected operators. In LQCD this is equivalent to reducing the variance of the trace of the inverse of the quark matrix. These methods reduce the computational needs as well because less matrix inversions are required to achieve the desired variance. 

Given two matrices $Q$ and $\tilde{Q}$, where $\tilde{Q}$ is traceless, we can show that expectation value of the trace of a matrix remains unchanged after adding a traceless matrix to it, i.e., the relationship 
\begin{equation}\label{xyz267}
\begin{aligned}
 \langle Tr (QX)\rangle =\Big \langle Tr \left\{ \left(Q-\tilde Q \right) X\right\}\Big \rangle
\end{aligned}
\end{equation}
holds true for a large number of noises. Note that the average is taken only over noises and that $ \langle X_{ij} \rangle = \delta_{ij} $.

We just observed that the trace remains invariant under addition of a traceless matrix. However, if we take the variance of the trace the relationship doesn't hold true. We can see from Eq.~(\ref{xyz255}) that variance depends on the off-diagonal elements of a matrix. So, if we can find a matrix $\tilde{Q}$ which is traceless but has the same off-diagonal elements same as $Q$, the new matrix $Q-\tilde{Q}$ is a diagonal matrix with zero variance. This can be understood from the relation 
\begin{equation}\label{xyz275}
    \begin{aligned}
    V\bigg[Tr \left\{ (Q-\tilde Q)X_{z(N\ge3)}\right\} \bigg] = \frac{1}{N} \sum_{i\not=j} \left(|q_{ij}-{\tilde q_{ij}}|^2 \right).
 \end{aligned}
\end{equation}
However, this situation is ideal. In practical applications it is difficult to mimic the off-diagonal elements of $Q$. 

In the case of LQCD calculations, $Q$ is actually the inverse of the quark matrix $M$. Therefore our goal is to find a matrix $\tilde{M}^{-1}$ that is traceless and has off-diagonal elements as close to $M^{-1}$ as possible. Besides the standard method of perturbative noise subtraction, we use five different techniques to create $\tilde{M}^{-1}$ which will be described in the following section.

\section{Subtraction Methods}\label{Sfive}

We have applied several new techniques to non-subtracted (NS) lattice data. These are termed Eigenvalue Subtraction (ES)\cite{vicpaper}, Hermitian Forced Eigenvalue Subtraction (HFES)\cite{Guerrero2} and Polynomial Subtraction (POLY)\cite{Liu2}. In \cite{vicpaper} we also introduced techniques which combine deflation with other subtraction methods. The method which combines HFES and POLY will be termed HFPOLY and the method which combines HFES and perturbative subtraction will be termed HFPS.
 
In LQCD the expectation value of any operator can be written as
\begin{equation}\label{xyz294}
\langle\bar{\psi}\Theta\psi\rangle=-Tr(\Theta M^{-1}).
\end{equation}
For the scalar operator, $\Theta$ can be replaced by the identity operator and the expectation value can be written as
\begin{equation}\label{xyz298}
\langle\bar{\psi}(x)\psi(x)\rangle=-Tr(M^{-1}).
\end{equation}
Using Eq.~(\ref{xyz168}), the trace of the inverse can be expressed as
\begin{equation}
\begin{aligned}
Tr(M^{-1}) &= \sum_i\langle x_i\eta_i\rangle\\&
=\sum_i \frac{1}{N}\sum_n^N x_i^{(n)}\eta_i^{*(n)}\\&
=\sum_i \frac{1}{N}\sum_n^N \left( \sum_j M_{ij}^{-1}\eta_j^{(n)}\right)\eta_i^{*(n)}\\&
=\sum_i \frac{1}{N}\sum_n^N \eta_i^{*(n)}\sum_j M_{ij}^{-1}\eta_j^{(n)}.
\end{aligned}
\end{equation}


The numerical, unmodified trace using this equation introduces a noise variance in the calculation which we wish to improve upon. Following the ideas of the last section, this can be done by creating a traceless matrix $\tilde {M}^{-1}$ that subtracts off the off-diagonal elements of $M^{-1}$. Addition of a traceless matrix $\tilde {M}^{-1}$  will now modify the relationship as
\begin{equation}
\begin{aligned}
Tr(M^{-1}) &=\sum_i \frac{1}{N}\sum_n^N \eta_i^{*(n)}\sum_j\left( M_{ij}^{-1}-\tilde{M}_{ij}^{-1}\right)\eta_j^{(n)}.
\end{aligned}
\end{equation}
In LQCD simulations, it is very difficult to generate traceless matrices $\tilde M^{-1}$. So, for the correct evaluation of the unbiased evaluation of the scalar operator, $Tr(\tilde M^{-1})$ needs to be added to the calculation. This gives
\begin{equation}
\begin{aligned}
Tr(M^{-1})
=\sum_i \frac{1}{N}\sum_n^N \eta_i^{*(n)}\sum_j\left( M_{ij}^{-1}-\tilde{M}_{ij}^{-1}\right)\eta_j^{(n)}+ Tr(\tilde M^{-1}).
\end{aligned}
\end{equation}
Simplification of this equation yields,
 \begin{equation}
Tr\left(M^{-1}\right)=\frac{1}{N}\sum_n^N{\left( \eta ^{(n)\dagger} \left(  x^{(n)} -\tilde x^{(n)}\right) \right)} + Tr\left(  \tilde M^{-1}\right),
\end{equation}
where $x^{(n)}$ is the solution vector generated when implementing the GMRES algorithms and $\tilde x^{(n)}$ is given by
\begin{equation}
\tilde x^{(n)}\equiv \tilde M^{-1}\eta^{(n)}.
\end{equation}
For any operator $\Theta$, the appropriate trace becomes
\begin{equation}
Tr\left( \Theta M^{-1}\right)=\frac{1}{N}\sum_n^N{\left( \eta ^{(n)\dagger} \Theta \left(x^{(n)} -\tilde x^{(n)}\right) \right)} + Tr\left( \Theta \tilde M^{-1}\right).
\end{equation}
We will see that the calculation of the corrected trace will not add significant computational time to the simulation. Also note that adding $Tr\left(\Theta  \tilde M^{-1}\right)$ has no influence on the noise error bar, which is what is studied here. We will now explain different ways of obtaining the subtraction matrix $\tilde{M}^{-1}$.

\subsection{Perturbative Subtraction (PS)}

Perturbative noise subtraction is the standard method used by researchers to estimate off-diagonal elements of the quark matrix. In this subsection we will explain how it does the estimation and in what regard it differs between local and non-local operators. 

For Wilson case, the quark matrix $M$ can be written as 
\begin{equation}\label{xyz556}
 \begin{aligned}
 M_{ij}=\delta_ {ij} -\kappa P_{ij},
 \end{aligned}
\end{equation}
where $i,j$ covers space-time-color-Dirac space and $P_{ij}$ is 
\begin{equation}
 \begin{aligned}
 P_{ij}=\sum_\mu \left[(1-\gamma_\mu) U_\mu (x) \delta_{{x,y}-a_\mu}+ (1+\gamma_\mu) U^\dagger_\mu (x-a_\mu) \delta_{{x,y}+a_\mu}\right].
 \end{aligned}
\end{equation}
If we take inverse of Eq.~(\ref{xyz556}), we get
\begin{equation}
 \begin{aligned}
 M_{ij}^{-1}= \frac{1}{\delta_{ij}-\kappa P_{ij}}.
  \end{aligned}
\end{equation}
We now can expand this equation in Taylor series for small values of $\kappa P$ to get the matrix $\tilde{M}^{-1}_{pert}$ such that
\begin{equation}
 \begin{aligned}
 \tilde M ^{-1}_{pert}= I +\kappa P + (\kappa P)^2 + (\kappa P)^3 + (\kappa P)^4 +......
\end{aligned}
\end{equation}
For simulations, we cannot calculate all the orders of $ \tilde M ^{-1}_{pert}$ and will have to settle for a finite order. Therefore $Tr\left(\tilde M ^{-1}_{pert}\right)$ will have to be added back for the correct trace computation. In this work the expansion is calculated up to seventh order of terms, i.e,
\begin{equation}
 \begin{aligned}
 \tilde M ^{-1}_{pert}= I +\kappa P + (\kappa P)^2 + (\kappa P)^3 + (\kappa P)^4 +  (\kappa P)^5 +  (\kappa P)^6.
\end{aligned}
\end{equation}
One then has
\begin {equation}\label{xyz601}
Tr\left( M^{-1}\right)=\frac{1}{N}\sum_n^N{\left( \eta ^{(n)\dagger}  \left(  x^{(n)} -\tilde x^{(n)}_{pert}\right) \right)} + Tr\left(  \tilde M^{-1}_{pert}\right).
\end {equation}
Generalizing for any operator $\Theta$, the trace now takes the form,
\begin {equation}\label{expV1}
Tr\left( \Theta M^{-1}\right)=\frac{1}{N}\sum_n^N{\left( \eta ^{(n)\dagger} \Theta \left(  x^{(n)} -\tilde x^{(n)}_{pert}\right) \right)} + Tr\left( \Theta \tilde M^{-1}_{pert}\right),
\end {equation}
where
\begin{equation}
\tilde x^{(n)}_{pert}\equiv\tilde M^{-1}_{pert}\eta^{(n)}.
 \end{equation}

The $\tilde M^{-1}_{pert}$ involved is not traceless so the diagonal elements of the matrix $M$ will be changed. This will therefore change the vacuum expectation value and we need to add some correction terms after the subtraction. Notice that only closed loop, gauge invariant objects contribute to the trace in Eq.(\ref{expV1}). In other words, only closed path objects contribute to the trace. 

\begin{figure}
 \begin{center}
\includegraphics[scale=.4]{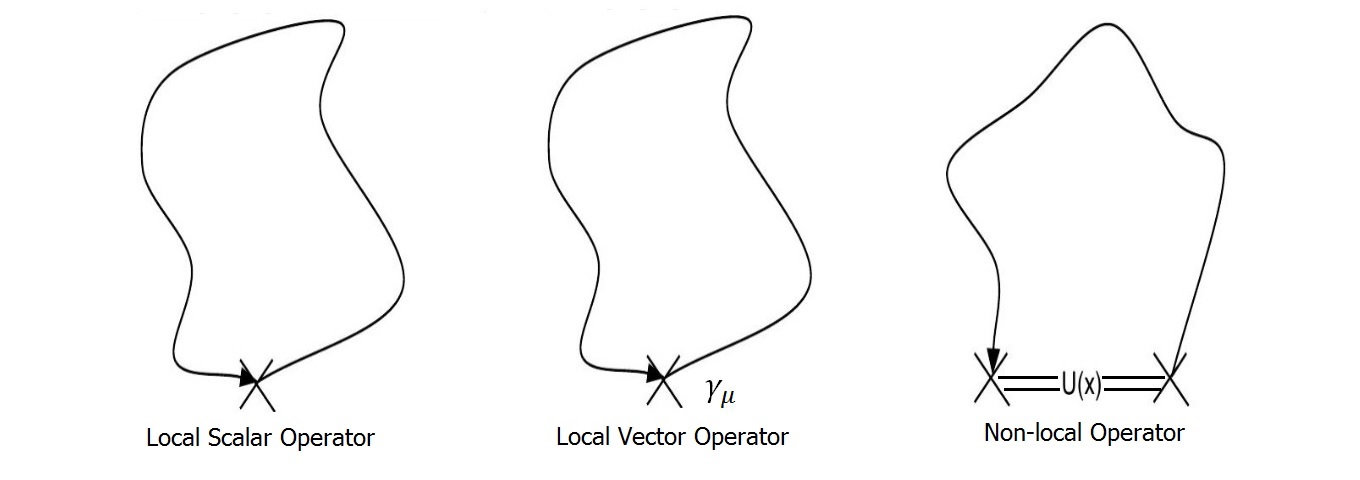}
 \end{center}
 \caption{General diagram of the quark line contributions for local scalar, local vector and non-local operators.}
 \label{fig:general}
\end{figure}

\begin{figure}[t!]
\centering
\includegraphics[trim={0 6cm 0 3cm},clip,width=1.25\textwidth]{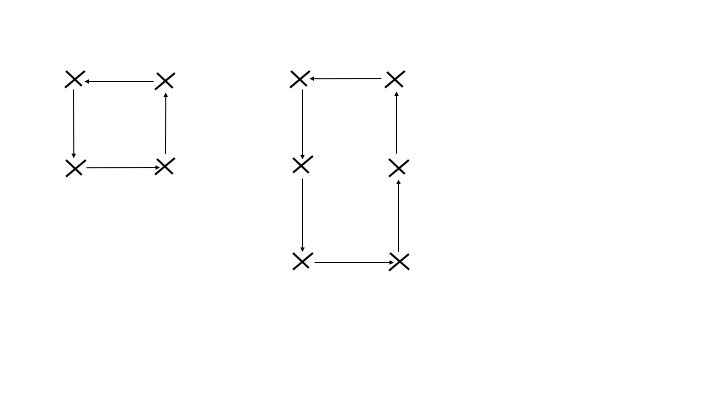}
\caption{Perturbative local operator contribution at 0($\kappa^4$) and 0($\kappa^6$).}
\label{pertur2}
\end{figure}

\begin{figure}[!htpb]
\centering
\includegraphics[trim={0 6cm 2cm 1cm},clip,width=1.25\textwidth]{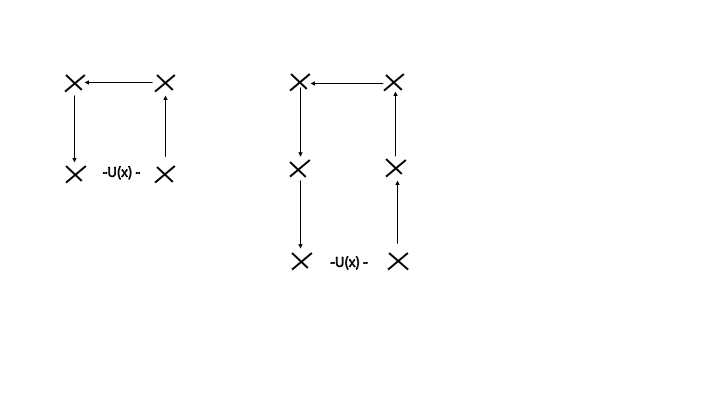}
\caption{Perturbative vector operator contribution at 0($\kappa^3$) and 0($\kappa^5$).}
\label{pertur3}
\end{figure}

The general picture of the local scalar, local vector and non-local operator is given in Fig.~\ref{fig:general}. 
The geometric interpretation of the perturbative expansion in Figs.~\ref{pertur2} and~\ref{pertur3} shows how each order of $\kappa$ is related to a path. We can see from the Fig.~\ref{pertur2} that local operators require a correction starting at $4^{th}$ order of $\kappa$ and from Fig.~\ref{pertur3} that point-split operators, which are spread across one lattice link, require a correction starting at $3^{rd}$ order of $\kappa$. Note that point-split operators have an inbuilt order of $\kappa$. Also, the local scalar operator requires a correction starting at $0^{th} $ order of $\kappa$. In general, only even orders of $\kappa $ contribute to the local operators and odd orders contribute to the non-local operators.

The direct calculation of the $Tr\left(  \Theta M^{-1}_{pert}  \right)$ is very expensive, however closed loops can be easily calculated\cite{Liu2,Guerrero2}. This can be done by solving the system,
\begin{equation}
 \begin{aligned}
\tilde{M}_{pert} {x}=e_i,
 \end{aligned}
\end{equation}
which does not involve a noise vector. $e_i$ is the unit vector in the $i^{th}$ direction that spans space-time-color-Dirac space. Now, we can solve for the solution vector by calculating
\begin{equation}
 \begin{aligned}
 x=\tilde M ^{-1}_{pert}e_i=\left( I + \kappa P + (\kappa P)^2 + (\kappa P)^3 + ....+(\kappa P)^n \right)e_i,
\end{aligned}
\end{equation}
in an iterative manner. Not all the $O(\kappa^{n})$ terms contribute to the calculation of trace for each specific operator. We drop the terms which do not contribute to the trace. That means for the calculation of non-local operators we drop the even orders of $\kappa$ while odd orders of $\kappa$ are dropped for calculation of local scalar and local vector operators. 
The trace of the corrected part is now calculated as 
\begin{equation}
Tr\left( \Theta \tilde{M}^{-1}_{pert}\right) = \sum_i\left( e_i \Theta \tilde{M}^{-1}_{pert}e_i\right),
\end{equation}
for local operators, and
\begin{equation}
Tr\left( \Theta (\mu)\tilde{M}^{-1}_{pert}\right) = \sum_i\left( e_{i+\mu} \Theta(\mu) \tilde{M}^{-1}_{pert}e_i\right),
\end{equation}
for a point-split (or nonlocal) operator $\Theta (\mu)$ pointing in the $\mu$ direction.

 \subsection{Polynomial Subtraction (POLY)}

The perturbative noise subtraction method is widely used not only because it reduces the error associated with LQCD simulations but also because $\tilde{M}^{-1}_{pert}$ is relatively easy to build. This fact motivated our group to develop a new technique called polynomial noise subtraction, also denoted as POLY\cite{Liu2}. The only difference is that the coefficients of the new matrix which mimics the off-diagonal elements are now different from one. i.e., $\tilde{M}^{-1}_{pert}$ which was initially represented to the 7th order as
\begin{equation}
\tilde{M}_{pert}^{-1} \equiv 1+\kappa P+(\kappa P)^2+(\kappa P)^3+(\kappa P)^4+(\kappa P)^5+(\kappa P)^6.
\end{equation}
is now replaced by $\tilde{M}^{-1}_{poly}$ such that
\begin{equation}
\tilde{M}_{poly}^{-1} \equiv a_1+ a_2\kappa P+a_3(\kappa P)^2+a_4(\kappa P)^3+a_5(\kappa P)^4+a_6(\kappa P)^5+a_7(\kappa P)^6,
\end{equation}
where the $a_i$'s are the coefficients obtained using MINimal RESidual (Min-Res) polynomial\cite{Liu2}. Let us explain how we can obtain these coefficients.

The basic linear equation system we discuss is 
\begin{equation}
Mx=\eta.
\end{equation}
Associated with this, we want to create a Krylov subspace $\mathcal{K}_t $ spanned by $b,Mb,M^2b,M^3b,....$ such that
\begin{equation}
\mathcal{K}_t = \{ b,Mb,M^2b,M^3b,... \}.
\end{equation}
Any solution vector $x_t$ can also be expressed  as a linear combination of elements from this subspace. i.e,
\begin{equation}
\begin{aligned}
x_t &= a_0b+a_1Mb+a_2M^2b+a_3M^3b+...\\&
=\left(  a_0+a_1M+a_2M^2+a_3M^3+...  \right)b \\&
=P\left(M\right)b,
\end{aligned}
\end{equation}
 where $P(M)$ is a polynomial of $M$. 
We want to mininize norm of the residual ${||Mx_t -b||}_2$  which now equals ${||(MP(M) -I)b||}_2$. We can see that after the residual is minimized, we obtain a polynomial which plays the role of the inverse quark matrix. This can be understood from the approximations
\begin{equation}\label{xyz634}
\begin{aligned}
&\quad\qquad{||Mx_t -b||}_2 \approx 0 \\&
\implies  {||(MP(M) -I)b||}_2 \approx 0 \\ &
\implies P(M) \approx M^{-1} .
\end{aligned}
\end{equation}

In the POLY method, we use the polynomial $P(M)$ obtained from the Min-Res projection to mimic off-diagonal elements of the inverse of a quark matrix. As we will see this method produces smaller variance than the PS method without a significant increase of matrix vector products. We will limit our investigation here to a 7th order polynomial (6th order in $\kappa P$). However, the polynomial can easily be extended to higher powers, where it becomes significantly more effective.

\begin{figure}[!htpb]
\centering
\includegraphics[width=.75\textwidth]{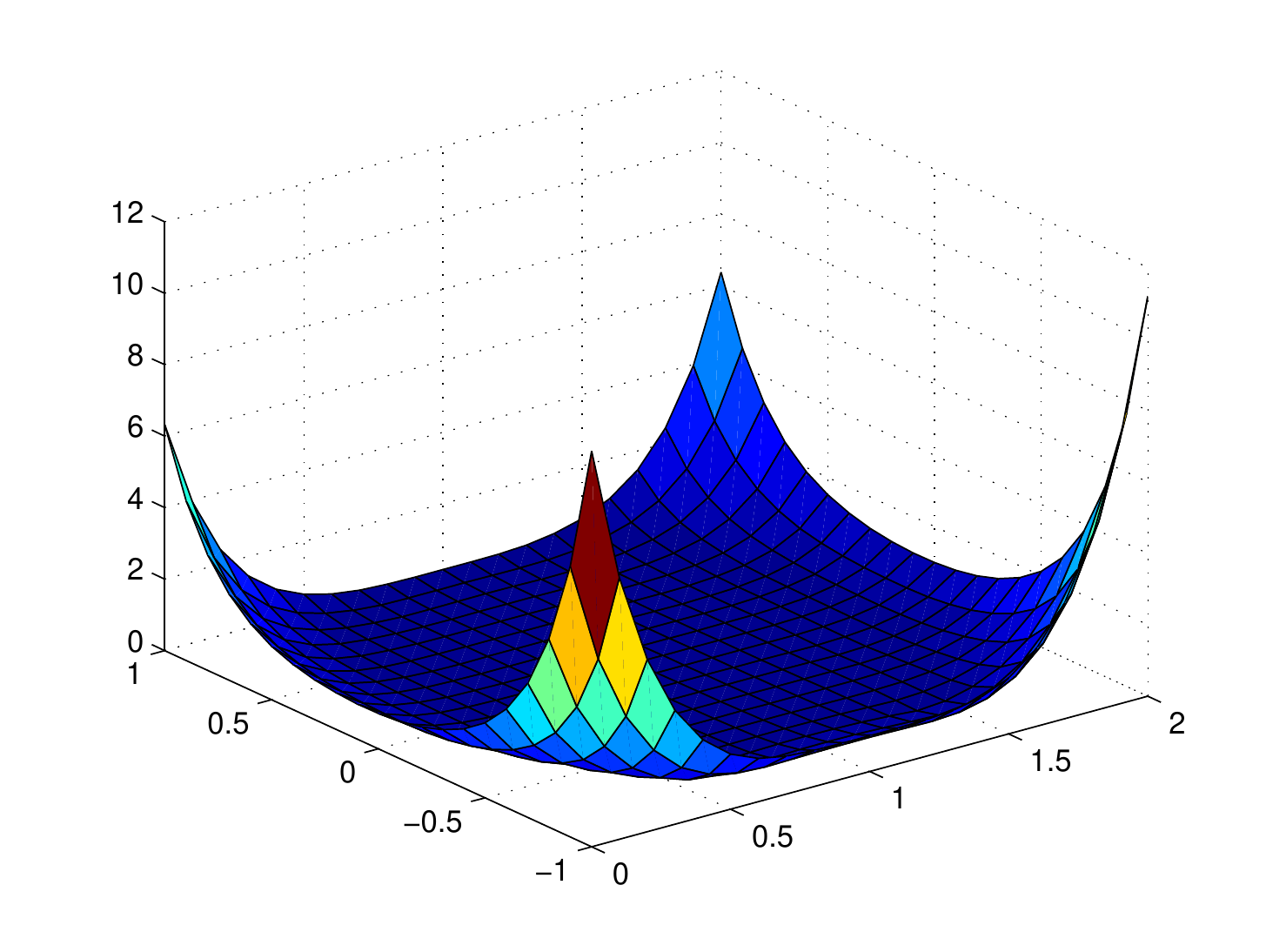}
\caption{The shape of seventh order MINRES polynomial.}
\label{minres}
\end{figure}

Consider a $(n+1)^{th}$  order polynomial of M such that
\begin{equation}\label{xyz646}
P_n(M) = a_0+a_1M+a_2M^2+....+a_nM^n,
\end{equation}
where the coefficients 
\begin{equation}
\vec{a}=\{ a_0,a_1,....a_n\}
\end{equation}
are obtained from Min-Res projection. By solving a set of equation 
\begin{equation}
\left(v^{\dagger}v\right) \vec{a} = v^\dagger b,
\end{equation}
where $v$ is a matrix given by 
\begin{equation}
v=\bigg( Mb\quad M^2b\quad....\quad M^{n+1}b\bigg),
\end{equation}
we can obtain the coefficients and build the polynomial $P(M)$. This polynomial is used to construct $\tilde{M}^{-1}_{poly}$, which is used exactly as $\tilde{M}^{-1}_{pert}$ in the PS method. We find that coefficients only need to be calculated for one single noise vector and then used for all the right hand sides. The coefficients of the $7^{th}$ order Min-Res polynomial for a $8^4$ lattice at $\kappa=0.157$ are listed in Table~\ref{xyz520}. From the table we see that the signs of consecutive coefficients are alternating. Also, these coefficients have a larger value for middle order coefficients and are far different from unity, as seen in the PS method. We can see from Fig.~\ref{minres} that the polynomial is a good approximation to $M^{-1}$ since it remains flat at $z=1$ plane in some region of the complex space\cite{Liu2}.
\begin{table}[ht!]
\centering
\begin{tabular}{ |c|c|c|} 
\hline
 Coefficients  &   real part  &    imaginary part\\ 
\hline
 $a_0$ &   6.4551 &    0.0263 \\
\hline 
 $a_1$ &  -20.5896 &  -0.1429  \\ 
 \hline
 $a_2$ &   39.3584 &  0.3563\\
\hline 
 $a_3$ &  -47.6275 &  -0.5076 \\ 
\hline 
 $a_4$ & 36.7205  &   0.4378\\ 
 \hline
 $a_5$&  -17.4702 & - 0.2265\\
\hline 
 $a_6$ & 4.6727 &  0.0648 \\ 
 \hline
$a_7$ &-0.5375&- 0.0079\\
\hline
 \end{tabular}
 \caption{Coefficients of the Min-Res polynomial.}
\label{xyz520}
\end{table}

Since $M=I+\kappa P$, we can express $\tilde{M}^{-1}_{poly}$ in terms of P as shown below

\begin{equation}
\begin{aligned}
\tilde{M}^{-1}_{poly} &= a_0+a_1M+a_2M^2+....+a_nM^n \\&
= a_0+a_1\left(I+\kappa P\right)+a_2\left(I+\kappa P\right)^2+....+a_n\left(I+\kappa P\right)^n \\&
= b_0+b_1\kappa P+b_2\kappa^2 P^2+....+b_n\kappa^n P^n.
\end{aligned}
\end{equation}
These coefficients $\{b_0,b_1,....b_n\}$ can easily be calculated from the coefficients $\{ a_0,a_1,....a_n\}$. We can build $\tilde{M}^{-1}_{poly}$ in the same way as we did for $\tilde{M}^{-1}_{pert}$ by dropping the terms that don't contribute to trace calculation.

The correction term for this method is 
\begin{equation}
Tr\left( O\tilde{M}^{-1}_{poly}\right) = \sum_i\left( e^*_i O \tilde{M}^{-1}_{poly}e_i\right),
\end{equation}
for local operators and
\begin{equation}
Tr\left( \Theta(\mu) \tilde{M}^{-1}_{poly}\right) = \sum_i\left( e_{i+\mu} \Theta(\mu) \tilde{M}^{-1}_{poly}e_i\right),
\end{equation}
for point-split. The trace in this method takes the form
\begin {equation}
Tr\left( \Theta M^{-1}\right)=\frac{1}{N}\sum_n^N{\left( \eta ^{(n)\dagger} \Theta \left(  x^{(n)} -\tilde x^{(n)}_{poly}\right) \right)} + Tr\left( \Theta \tilde M^{-1}_{poly}\right),
\end {equation}
where
\begin{equation}
\tilde x^{(n)}_{poly}\equiv\tilde M^{-1}_{poly}\eta^{(n)}.
 \end{equation}
 
\subsection{Eigenvalue Subtraction (ES)}

PS and POLY both involve an expansion in terms of $\kappa$. Such methods lose their effectiveness at realistically small quark mass values, which for the Wilson action means near $\kappa_{crit}$. On the other hand, one would expect deflation methods, which effectively remove small eigenvalues from quantities, to become more effective in this limit. Eigenvalue subtraction (ES) is one such method developed by our group to cope with this problem. Although this technique is extremely straightforward, we will see that it gets flooded by highly non-normal eigenvalues from the nonhermitian Wilson action and often increases the variance\cite{Baral,Gambhir}. However, it paves a path to develop better methods which we will describe shortly.

The spectrum of low eigenvalues of matrices can limit the performance of iterative solvers. We have emphasized the role of deflation in accelerating the convergence of algorithms in Ref.~\cite{Wilcox2}. Here we investigate deflation effects in statistical error reduction. Consider the vectors $e_R^{(q)}$ and $e_L^{(q)\dagger}$, which are defined as normalized right and left eigenvectors of the matrix $M$, as in
\begin{equation}
Me_R^{(q)}=\lambda^{(q)} e_R^{(q)},
\end{equation}
and
\begin{equation}
 e_L^{(q)\dagger} M=\lambda^{(q)} e_L^{(q)\dagger},
\end{equation}
where $\lambda^{(q)}$ is the eigenvalue associated with both eigenvectors. One can show the orthonormal relationship for all eigenvectors is
\begin{equation}
e_L^{(q)\dagger}e_R^{(q')}=\delta_{q\,q'}.
\end{equation}
With a full set $N$ of eigenvectors and eigenvalues the matrix $M$ can be fully formed as 
\begin{equation}
M=\sum_{q=1}^{N} {e_R^{(q)}  \lambda^{(q)} e_{L}^{(q)\dagger}},
\end{equation}
or
\begin{equation}
M=V_R \Lambda V_L^{\dagger}, 
\end{equation}
where $V_R$ contains the right eigenvectors and $V_L^{\dagger}$ contains the left eigenvectors. $\Lambda$ is a purely diagonal matrix made up of the eigenvalues of $M$ in the order they appear in both $V_R$ and $V_L$. 

The ES method involves forming a matrix, $\tilde{M}^{-1}$ by utilizing eigenspectrum decomposition. It is computationally unfeasible to use all the eigenmodes of $M$, which means we can form only an approximation to ${M}^{-1}$ by using only the $Q$ smallest pairs where $Q$ is determined based on the computational resources. We name the subtraction matrix thus formed as $\tilde{M}^{-1}_{eig}$.

Deflating out eigenvalues with the linear equation solver GMRES-DR can mimic the low eigenvalue structure of the inverse of matrix $M$ as
\begin{equation} \label{xyz640}
\begin{aligned}
\tilde{M}_{eig}^{-1} \equiv \tilde{V}_{R} \tilde{\Lambda}^{-1} \tilde{V}_{L}^{\dagger},
\end{aligned}
\end{equation}
where $\tilde{V}_{R}$ and  $\tilde{V}_{L}^{\dagger}$ are the computed right and left eigenvectors and $ \tilde{\Lambda}^{-1}$ is the inverse of eigenvalues. Here, $\tilde{V}_R$ is $\mathcal{N}\times Q$, $\tilde{\Lambda}^{-1}$ is size $Q\times Q$ and $\tilde{V}_{L}^{\dagger}$ is of size $Q \times \mathcal{N}$ which makes $\tilde{M}_{eig}^{-1}$ same size as ${M}^{-1}$.

In index form, Eq.~(\ref{xyz640}) can be written as
\begin{equation}
 \begin{aligned}
\left[\tilde M_{eig}^{-1}\right]_{ij} = \sum_q^{Q}  \sum_k^{Q} \left({\tilde V_{R}}\right)_{ik} {\tilde \Lambda_{kq}^{-1}} \left( \tilde V_{L}^{\dagger}\right)_{jq} .
  \end{aligned}
\end{equation}
We can further simplify it using $ {\tilde \Lambda_{kq}^{-1}} =  \lambda_k^{-1} \delta _{kq}$ such that
\begin{equation}
 \begin{aligned}
 {\left[ \tilde {M}_{eig}^{-1}\right]}_{ij}  &= \sum_q^{Q} \left( \sum_k^{Q} \left({\tilde{V}_{R}}\right)_{ik} \lambda_k^{-1} \delta _{kq}\right) \left( \tilde V_{L}^{\dagger}\right)_{jq} \\ &
 =\sum_q^{Q} (e^{(q)}_R)_i\,\lambda_q^{-1} (e^{(q)\dagger}_L)_j.
  \end{aligned}
\end{equation}

We choose the smallest $Q$ eigenvalues (in the complex plane) of the matrix $M$ to form $\tilde {M}_{eig}^{-1}$ because their contributions is greatest to the trace. Application of this eigenspectrum formulation enables us to effectively deflate out the off-diagonal elements for smaller quark masses. 

The vacuum expectation term $Tr\left( \tilde{M}^{-1}_{eig}\right)$ can not be calculated by analyzing the closed loops as we did in the PS and POLY methods. Instead, they have to be calculated directly as
\begin{equation}
 \begin{aligned}
 Tr(\tilde M _{eig}^{-1})= \sum_q^Q \frac{1}{\lambda^{(q)}}.
   \end{aligned}
\end{equation}
The trace of the inverse matrix now takes the form 
\begin {equation} 
Tr\left(  M^{-1}\right)=\frac{1}{N}\sum_n^N{\left( \eta ^{(n)\dagger}   \left(  x^{(n)} -\tilde x^{(n)}_{eig}\right) \right)} + Tr\left(  \tilde M^{-1}_{eig}\right),
\end {equation}
and
\begin {equation}
Tr\left( \Theta M^{-1}\right)=\frac{1}{N}\sum_n^N{\left( \eta ^{(n)\dagger}  \Theta \left(  x^{(n)} -\tilde x^{(n)}_{eig}\right) \right)} + Tr\left( \Theta \tilde M^{-1}_{eig}\right),
\end {equation}
where 
\begin{equation}
\begin{aligned}
\tilde x^{(n)}_{eig}& = \tilde M_{eig}^{-1}\eta^{(n)}\\&
= \sum_q^Q \frac{1}{\lambda_q}e_R^{(q)}\left( e_L^{(q)\dagger}\eta^{(n)} \right).
\end{aligned}
\end{equation}
This last operation does not add matrix vector products. The generation of eigenmodes only requires the super-convergence solution of a single right hand side with GMRES-DR. As pointed out previously, there is a relation between the even-odd eigenvectors for the reduced system and the full eigenvectors. Other right hand sides are accelerated with GMRES-Proj (similar algorithm projected over eigenvalues) using the eigenvalues generated.





\subsection{Hermitian Forced Eigenvalue Subtraction (HFES)}

To avoid the  non-normal eigenvalue issue in the nonhermitian case, we force our matrix to be formulated in a hermitian manner. The easiest way for us to do this with the Wilson matrix is to multiply by the Dirac $\gamma_5$ matrix. We can then form the low eigenvalue structure of $M\gamma_5$ from these eigenvalues. We define 
\begin{equation}
M'\equiv M\gamma_5.
\end{equation}
The inverse of this matrix is written as
\begin{equation}
{M'}^{-1}\equiv \gamma_5 M^{-1}.
\end{equation}
The trace of the original inverse matrix can be obtained from new matrix ${M'}$ without biasing the overall answer. From the relation
 \begin{equation}
 \begin{aligned}
 Tr(\gamma_5 M'^{-1}) &
 = Tr(\gamma_5 \gamma_5 M^{-1} ) \\&
 = Tr (M^{-1}),
  \end{aligned}
\end{equation}
we can see hermitian matrix $M'$ can be used to obtain the trace values associated with $M^{-1}$. It is important for the algorithm to do the multiplication on the {\it right} such that $M'=M\gamma_5$ to avoid using cyclic matrix properties which fail in finite noise space.

We can now form normalized eigenvectors ${e'}^{(n)}$, eigenvalues $\lambda '^{(n)}$ and solution vectors $\tilde{x}'{_{eig}^{(n)}}$ for this new hermitian matrix and perform a calculation similar to the ES method, accounting properly for the extra $\gamma_5$ factors. The trace of any operator  $\Theta$, for the HFES method takes the following form
\begin {equation}
Tr\left( \Theta M^{-1}\right)=\frac{1}{N}\sum_n^N{\left( \eta ^{(n)\dagger}  \Theta \left(  x^{(n)} -\tilde{x}'{_{eig}^{(n)}}\right) \right)} + Tr\left( \Theta \gamma_5\tilde {M}'{_{eig}^{-1}}\right),
\end {equation}
where
\begin{equation}
\tilde{x}'{_{eig}^{(n)}} \equiv\gamma_5\tilde{M}'{_{eig}^{-1}}\eta^{(n)}
                           =\gamma_5 \sum_{q}^{Q}{\frac{1}{{\lambda ' }^{(q)}}  { e'}^{(q)}  \left({e'}^{(q)\dagger} \eta^{(n)} \right) },
\end{equation}
and 
\begin{equation}
\tilde{M}'{_{eig}^{-1}} \equiv \tilde{V}' \tilde{\Lambda'}^{-1} \tilde{V}'^{\dagger}.
\end{equation}
$\tilde{V}'$ is a matrix whose columns are the $Q$ smallest right eigenvalues of $M'$. $\tilde{\Lambda '}^{-1}$ is the diagonal matrix of size $Q$ that contains the inverse of eigenvalues $1/{{\lambda'}^{(q)}}$ as the diagonal elements. The price paid here, similar to the ES method, is a single extra super-convergence on one right hand side for the hermitian system $M'$ with GMRES-DR or MINRES-DR to extract eigenvectors and eigenvalues.


\subsection{Combination Methods (HFPOLY and HFPS)}

We have also developed two methods which combine the error reduction techniques of HFES with POLY and PS, called HFPOLY and HFPS. N\"aively, for POLY we could think of this method as a subtracted combination: $\tilde M^{-1}_{poly}+\gamma_5 \tilde M'^{-1}_{eig}$. However, this presents a possible conflict since $\tilde M^{-1}_{poly}$ will overlap on the deflated hermitian eigenvector space. In order to prevent this, we also remove low eigenmode information from $\tilde {M}^{-1}_{poly}$. Since $\tilde {M}_{poly}$ is not hermitian, the procedure is to define $\tilde M'_{poly}=\tilde {M}_{poly}\gamma_5$ and remove its overlapping hermitian eigenvalue information using the eigenvectors from $M'$. Following the idea in Ref.~\cite{vicpaper}, we define
\begin{equation}
{e'}^{(q)\dagger}\tilde M'^{-1}_{poly}{e'}^{(q)}\equiv\frac{1}{\xi'^{(q)}},
\end{equation}
where ${e'}^{(q)}$ is the eigenmode of $M'$ generated within HFES method and $1/\xi'^{(q)}$ are the approximate eigenvalues of $\tilde M'^{-1}_{poly}$. The trace takes the following form,
 \begin{eqnarray}                                 
Tr\left(  \Theta M^{-1}\right ) = \frac{1}{N} & \sum_n^N { \left( \eta^{(n)\dagger} \left[  \Theta x^{{(n)}}  - \Theta\tilde{x}'{_{eig}^{(n)}} - \left( \Theta \tilde{x}_{poly}^{(n)}  - \Theta \tilde{x}'{_{eigpoly}^{(n)}} \right)  \right ] \right) } \nonumber\\+& Tr \left(  \Theta \gamma_5 \tilde{M'}_{eig}^{-1}\right)+Tr\left( \Theta \tilde{M}_{poly}^{-1}- \Theta \gamma_5 \tilde{M'}_{eigpoly}^{{-1}} \right) ,
\end{eqnarray}
where
 \begin{equation}
        \tilde{x}'{_{eigpoly}^{(n)}} \equiv\gamma_5\tilde{M}'{_{eigpoly}^{-1}}\eta^{(n)}
                           =\gamma_5 \sum_{q}^{Q}{\frac{1}{\xi'^{(q)}}  { e'}^{(q)}  \left({e'}^{(q)\dagger} \eta^{(n)} \right) }.
\end{equation}
$\tilde{x}'{_{eig}^{(n)}}$ and $\tilde x^{(n)}_{poly}$ are defined in previous sections and
\begin{equation}
\tilde{M}_{eigpoly}^{'^{-1}}  \equiv \tilde{V}' \Xi^{-1} \tilde{V}'^{\dagger},
\end{equation}
where $\tilde{V}'$ is defined above also and $\Xi^{-1}$ is the diagonal matrix of size $Q$ that contains approximate inverse eigenvalues, $1/\xi'^{(q)}$. 

In the case of HFPS, $\tilde{M}_{poly}^{-1}$ is replaced by $\tilde{M}_{pert}^{-1}$ and all the calculations are repeated.
     
\section{Operators under simulation}\label{Ssix}

The expectation value of any operator $\Theta$ is given as 
\begin{equation}\label{xyz286}
   \begin{aligned}
   \langle \bar {\psi}\Theta \psi  \rangle = -Tr(\Theta M^{-1}).
 \end{aligned}
\end{equation}
As pointed out above, we concentrate in this study on the lattice operators which respond best to unpartitioned or undiluted noise vectors. Also, even though each operator is calculated with a real and imaginary part, only the real or imaginary term contains a signal. This is because the quark propagator identity $S=\gamma_5 S^{\dagger}\gamma_5$, depending on the operator, allows only the real or imaginary part to be non-zero on a given configuration for each space-time point. 


We present the results for nine different operators which can be divided into three different groups. Allowing $\mu$ to range from 1 to 4, three groups transform into 9 different operators. Symbols and the equations related are shown in Table~\ref{Table903} below.
\begin{table}[ht!]
\centering
 \caption{Names of calculated operators and their field representations.}
\begin{tabular}{ c c c  } 
\\
\hline
 Name  & Representation  & Total operators \\ 
\hline
 Scalar & $Re[\bar{\psi}(x)\psi(x)]$ &  1  \\
 Local Vector & $Im[\bar{\psi}(x)\gamma_{\mu} \psi(x)] $ &  4 \\ 
 Point-Split Vector & $ \kappa  Im \big[\bar \psi(x+{a_\mu})  (1+\gamma_\mu)U^{\dagger}_\mu(x) \psi (x)\big]$  &  4\\ 
 &$ -\kappa Im \big[\bar\psi (x) (1-\gamma_\mu) U_\mu (x)\psi (x+ a_\mu)\big]$ & \\
\hline
 \end{tabular}
\label{Table903}
\end{table}

\section{Results}\label{Sseven}
The goal of this work is to develop new noise subtraction methods and to compare the statistical error bars generated from these methods to conventional noise subtraction techniques, in particular, perturbative subtraction (PS). In this section, we will compare numerical results for the statistical error bars from the various methods outlined in Section \ref{Sfive}, and show the effectiveness of our combination methods.

We will be discussing the results from two different lattice sizes, $8^4$ and $24^{3}\times 32$, using MATLAB and a modified version of Randy Lewis' FORTRAN 90 QQCD program. This was done to examine the effect of lattice size on our results as well as for debugging purposes. We work with the standard Wilson matrix in the quenched approximation at $\beta=6.0$ with periodic fermion boundary conditions. For the Wilson matrix, standard even/odd preconditioning was employed to reduce the computational time of the matrix inversion. This preconditioning creates a reduced system from the Wilson matrix, whose solution vectors and eigenmodes are easily related to the full Wilson matrix \cite{vicpaper,Guerrero2}. For the hermitian Wilson matrix, the full system was solved since there is no relation between even-odd eigenvalues and the full ones. In order to isolate the effect of decreasing quark mass, we use a common set of 200 $Z4$ noises for three different set of kappa values: $0.155, 0.156$ and $\kappa_{crit}=0.157$. Results at $\kappa_{crit}$ for the $24^3 \times32$ lattice are presented in Section \ref{Seight}. In order to guard against possible noise correlations, we also ran a sub-test on the $24^3 \times32$ lattice at $\kappa = 0.156$ by changing to a independent noise string. This test also gives an indication of the internal variance of the subtraction techniques in Table~\ref{1560large}. We found a small overall variability of $\sim10\%$ in the error bar results. Linear equations and eigenvalue equations  of the reduced Wilson matrix are solved using GMRES-DR($m$,$k$) for the first noise and GMRES-Proj for the remaining noises. The solution to the linear equations of the reduced Wilson matrix for the first noise vector was computed to super-convergence in order to obtain accurate eigenvalues and eigenvectors to be used in GMRES-Proj. The eigenvalues and eigenvectors of the full Wilson matrix were formed from those of the reduced system and used in the Eigenvalue Subtraction (ES) method. The linear equations of the hermitian Wilson matrix were also computed to super-convergence to obtain accurate eigenvalues and eigenvectors to be used in the hermitian subtraction methods. The size of the Krylov subspace in all these calculations was $m=200$, and the eigenmode space was $k=160$. The performance of the GMRES for the nonhermitian system was extremely efficient at all $\kappa$ values. However, it was necessary to switch to MINRES-DR for the hermitian system at $\kappa_{crit}$. More detials on the performance of the inversion algortithms will be presented in Section~\ref{Seight}.

Numerical calculations were carried out using both MATLAB and FORTRAN 90. Initial tests were performed in MATLAB on a lattice of size $8^4$ with the local vector and scalar operators to perform debugging and consistency checks. The point-split vector operator was not included in the initial tests in MATLAB due to the difficulty of its implementation. We then carried out calculations for a lattice of the same size in FORTRAN 90 for the local, scalar and point-split operators. 
After ensuring that the results were the same on both platforms, we proceeded to perform the simulations with a lattice of size $24^3 \times32$. On average, one out of every ten gauge configurations of size $8^4$ have two lowest lying eigenvalues that are drastically smaller than the next, especially so at $\kappa_{crit}$. Because of this, one hundred configurations were generated for a lattice of this size, and the statistical error was averaged over the number of configurations. For lattices of size $24^3 \times32$, the lowest lying eigenvalues do not vary as widely between configurations, so ten configurations were generated, and the error was averaged over. Our figures below show the statistical error as a function of the subtracted eigenmodes for a particular operator at a particular value of $\kappa$. 130 eigenvalues and eigenvectors across all 100 configurations were obtained from the hermitian Wilson matrix at $\kappa = 0.155$, and 140 across all 100 configurations were obtained for $\kappa = 0.156$ and 0.157, for the lattice of size $8^4$. 130 eigenvalues and eigenvectors across all ten configurations were obtained from the hermitian Wilson matrix at $\kappa = 0.155$  for the lattice of size $24^3 \times32$, and 149 eigenvalues and eigenvectors across all ten configurations were obtained at $\kappa = 0.156$ for the lattice of the same size. Included in each figure is the statistical error for no subtraction (NS) along with the error for the conventional PS method for reference. Table \ref{Table978} displays the legend for each figure. Results for the vector operators were obtained for all spacetime directions, but the results are independent of direction, so only the time component of each operator is presented.

\begin{table}[ht!]
\centering
 \caption {The legend on each graphs and their representation.} 
 \begin{tabular}{ c c c c } 
 \\
\hline
 Symbol  & Method  & Representation\\ 
 \hline
NS & No Subtraction &  Dark blue  \\
 ES & Eigenspectrum Subtraction  &  dotted green \\ 
 HFES & Hermitian Forced - Eigenspectrum Subtraction & Magneta\\
 PS & 7th Order Perturbative Subtraction &  Red\\ 
 POLY & 7th Order Polynomial Subtraction &  solid green \\ 
 HFPS & HFES and PS combination & Blue\\
 HFPOLY &HFES and POLY combination  & Pink\\ 
 \hline
 \end{tabular}
\label{Table978}
\end{table}

\begin{figure}[!htpb]
\centering
\includegraphics[width=.75\textwidth]{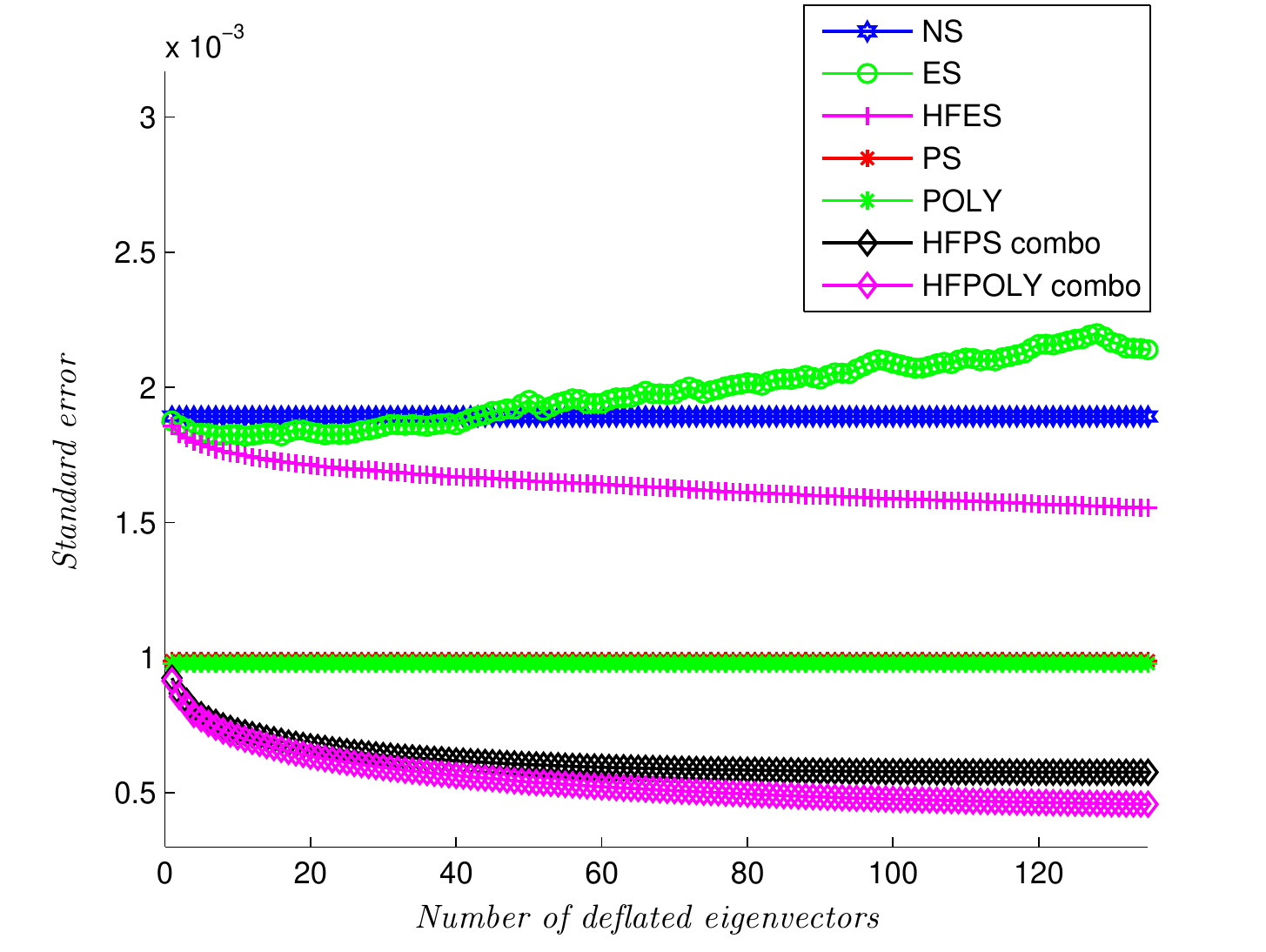}
\caption{Quenched $8^4$ lattice simulation results for the standard error as a function of deflated eigenvalues for the $J_4$ local operator at $\kappa = 0.155$.}
\label{figs54}
\end{figure}
\begin{figure}[!htpb]
\centering
\includegraphics[width=.75\textwidth]{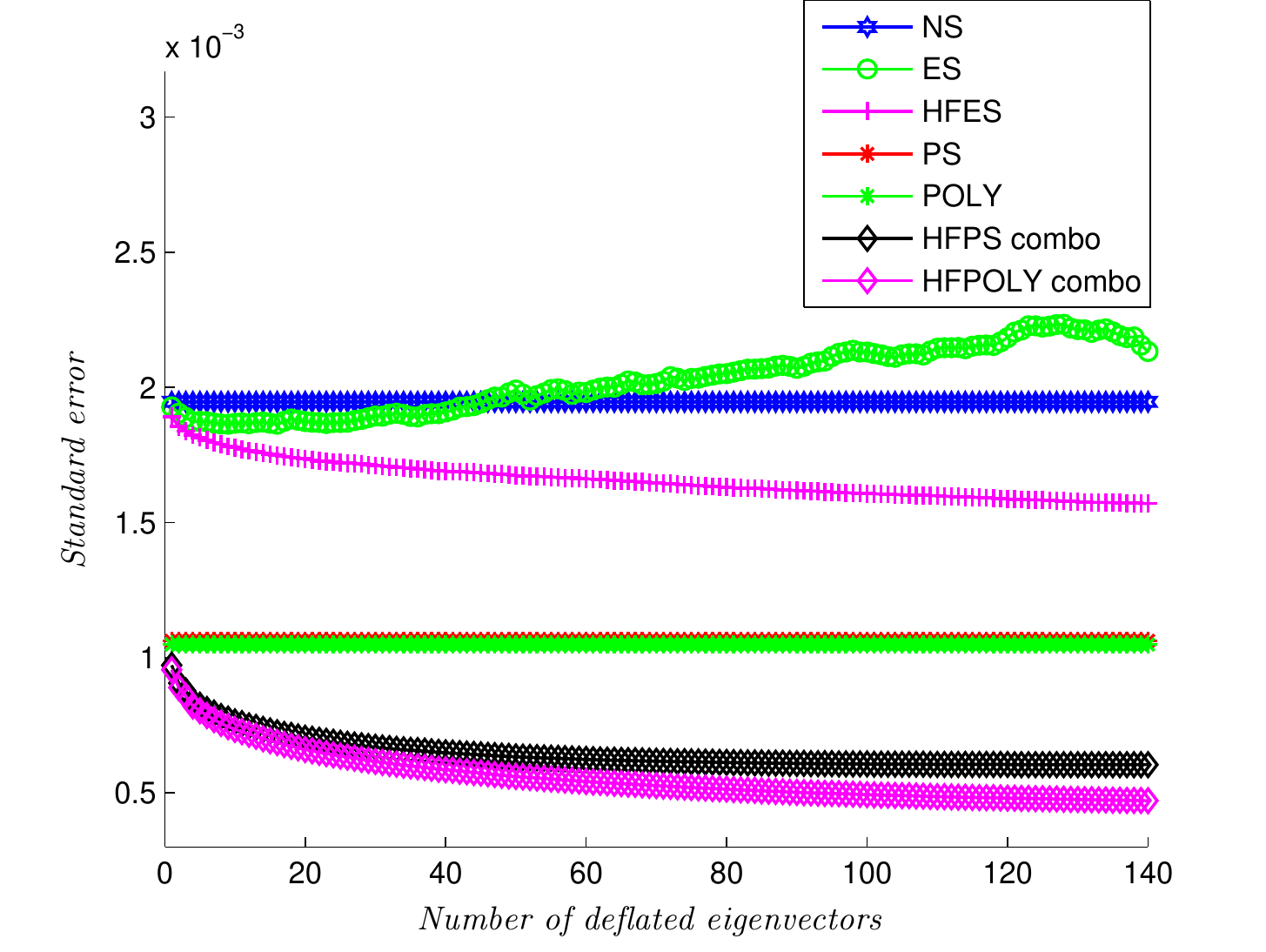}
\caption{Quenched $8^4$ lattice simulation results for the standard error as a function of deflated eigenvalues for the $J_4$ local operator at $\kappa = 0.156$.}
\label{figs64}
\end{figure}
\begin{figure}[!htpb]
\centering
\includegraphics[width=.75\textwidth]{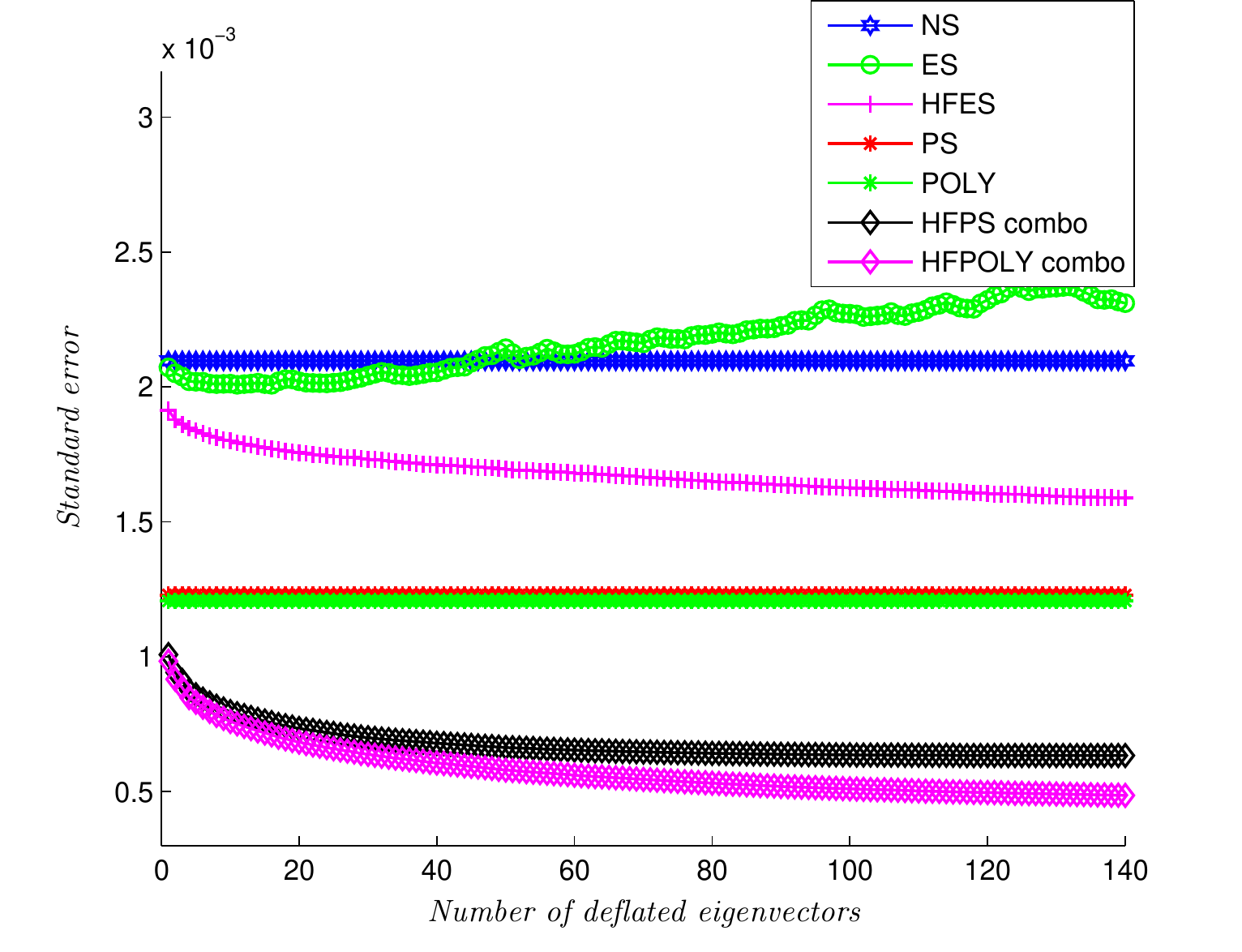}
\caption{Quenched $8^4$ lattice simulation results for the standard error as a function of deflated eigenvalues for the $J_4$ local operator at $\kappa = 0.157$.}
\label{figs74}
\end{figure}
\begin{figure}[!htpb]
\centering
\includegraphics[trim={1.1cm 8cm 0 8.5cm},clip,width=1.10\textwidth]{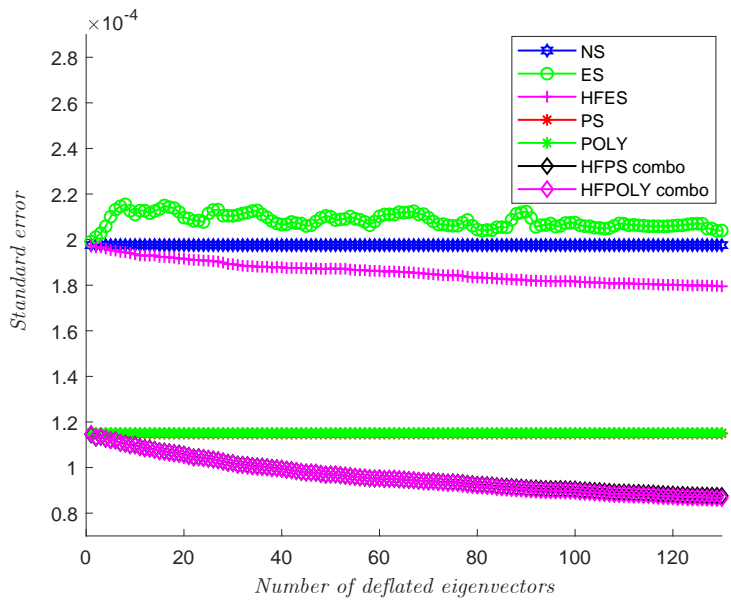}
\vspace{-.75cm}
\caption{Quenched $24^3\times 32$ lattice simulation results for the standard error as a function of deflated eigenvalues for the $J_4$ local operator at $\kappa = 0.155$.}
\label{figl54}
\end{figure}
\begin{figure}[!htpb]
\centering
\includegraphics[trim={1.1cm 8cm 0 8.5cm},clip,width=1.10\textwidth]{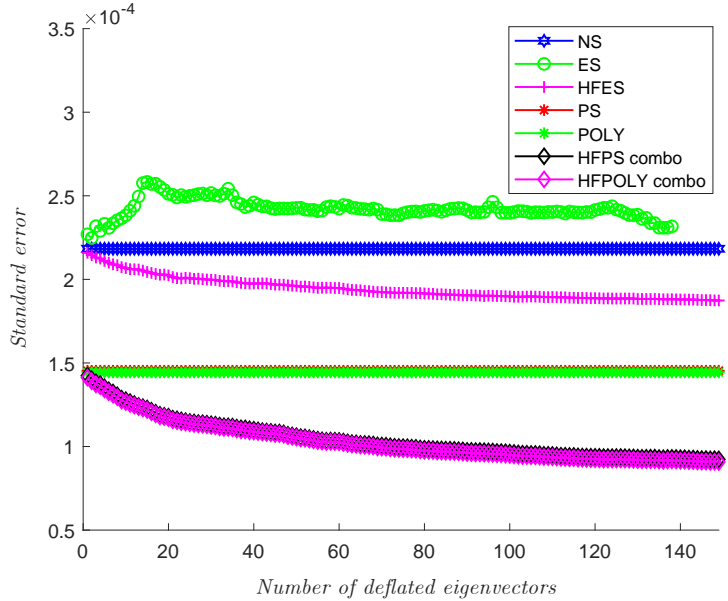}
\vspace{-.75cm}
\caption{Quenched $24^3\times 32$ lattice simulation results for the standard error as a function of deflated eigenvalues for the $J_4$ local operator at $\kappa = 0.156$.}
\label{figl64}
\end{figure}

Figs.~\ref{figs54},  \ref{figs64} and  \ref{figs74} display the statistical error for the $J_4$ local vector opertor, also called the charge density, for the $8^4$ lattice, with $\kappa = 0.155$, 0.156, and 0.157, respectively. Figs.~\ref{figl54} and \ref{figl64} display the statistical error for the $J_4$ local vector for the $24^3\times32$ lattice, at $\kappa = 0.155$ and 0.156, respectively.

The ES method produces error bars that are even greater than those produced from the NS method in all cases for this operator. This method shows no sign of decreasing the variance as the number of deflated eigenvalues increases. As understood in earlier work\cite{Guerrero2}, the ES method is inefficient because it can not produce orthogonal right eigenvectors for a nonhermtian system. Such eigenmodes overlap in right-eigenvector space and removing one can often enhance rather than decrease the variance.

In contrast, the HFES method reduces the error bars significantly in comparison to the NS method on the $8^4$ lattices. As the number of deflated eigenvectors is increased, the calculated error bars also decrease uniformly. If we compare the HFES results from Figs.~\ref{figs54},  \ref{figs64} and \ref{figs74}, one observes that the HFES error is continuous with the NS method for $\kappa=0.155$ at the lowest subtracted eigenvalue, while for  $\kappa = 0.156$ and 0.157 a discontinuity between the HFES and NS methods has developed. This occurred when the smallest eigenvalue was significantly smaller than second smallest eigenvalue. 


We observe similar improvement in the error calculations of this operator for the $24^3\times 32$ lattices with the HFES method. As $\kappa$ increases from 0.155 to 0.156, the separation between HFES and NS increases. However, HFES fails to decrease the error bars below those calculated via the PS method. Additionally, the POLY method is comparable to PS for both lattice volumes and both values of $\kappa$. The error bars generated through the HFPS and HFPOLY methods are also comparable to one another, and both decrease the error in comparison to PS for both lattice volumes and values of $\kappa$, with HFPOLY slightly more efficient than HFPS. As $\kappa$ increases, we observe an increased efficiency of the deflated HFPS and HFPOLY methods compared to PS for both the $8^4$ and the $24^3\times 32$ lattices. 

\begin{figure}[!htpb]
\centering
\includegraphics[width=.75\textwidth]{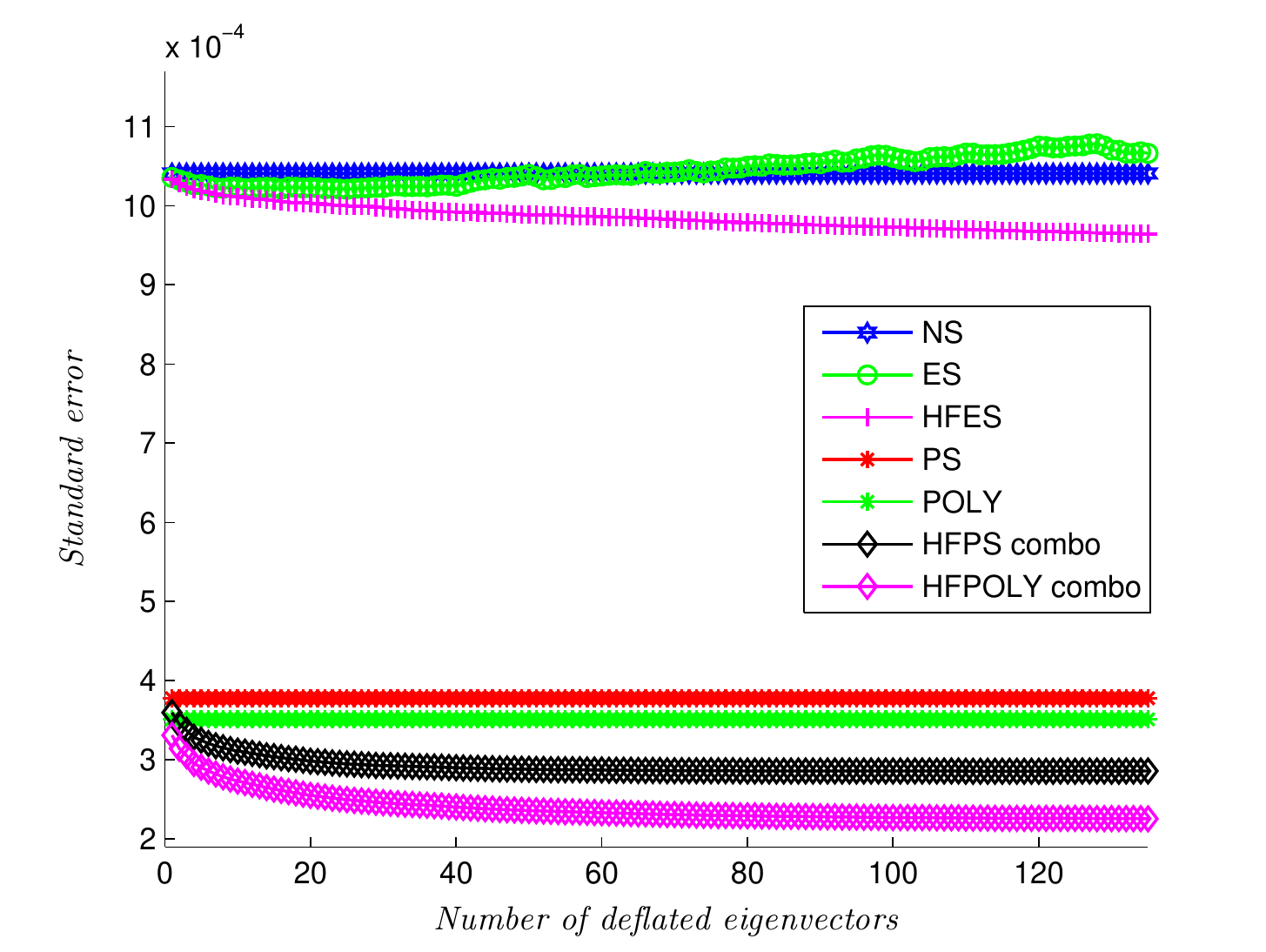}
\caption{Quenched $8^4$ lattice simulation results for the standard error as a function of deflated eigenvalues for the $J_4$ point-split operator at $\kappa = 0.155$.}
\label{figns54}
\end{figure}
\begin{figure}[!htpb]
\centering
\includegraphics[width=.75\textwidth]{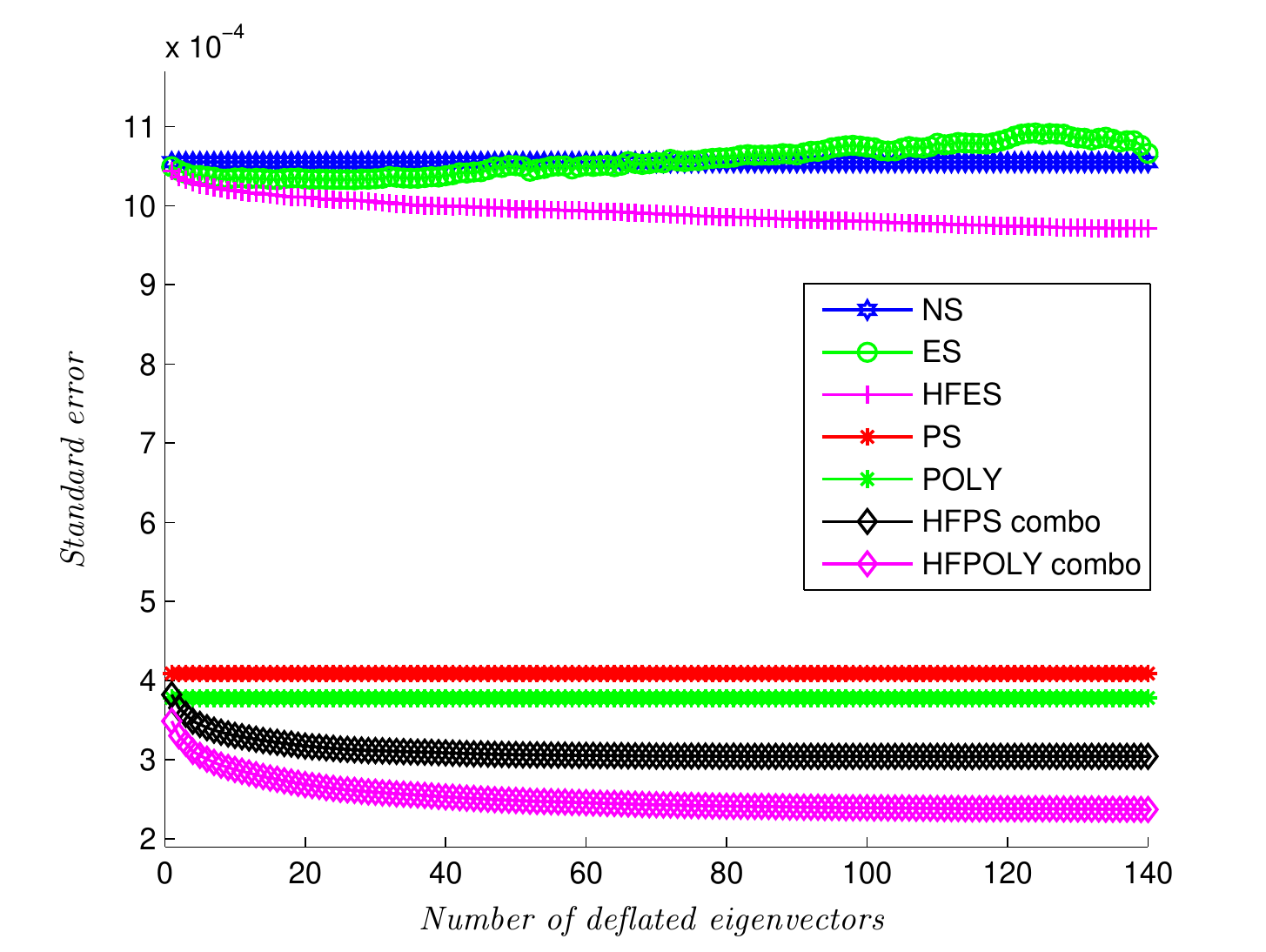}
\caption{Quenched $8^4$ lattice simulation results for the standard error as a function of deflated eigenvalues for the $J_4$ point-split operator at $\kappa = 0.156$.}
\label{figns64}
\end{figure}
\begin{figure}[!htpb]
\centering
\includegraphics[width=.75\textwidth]{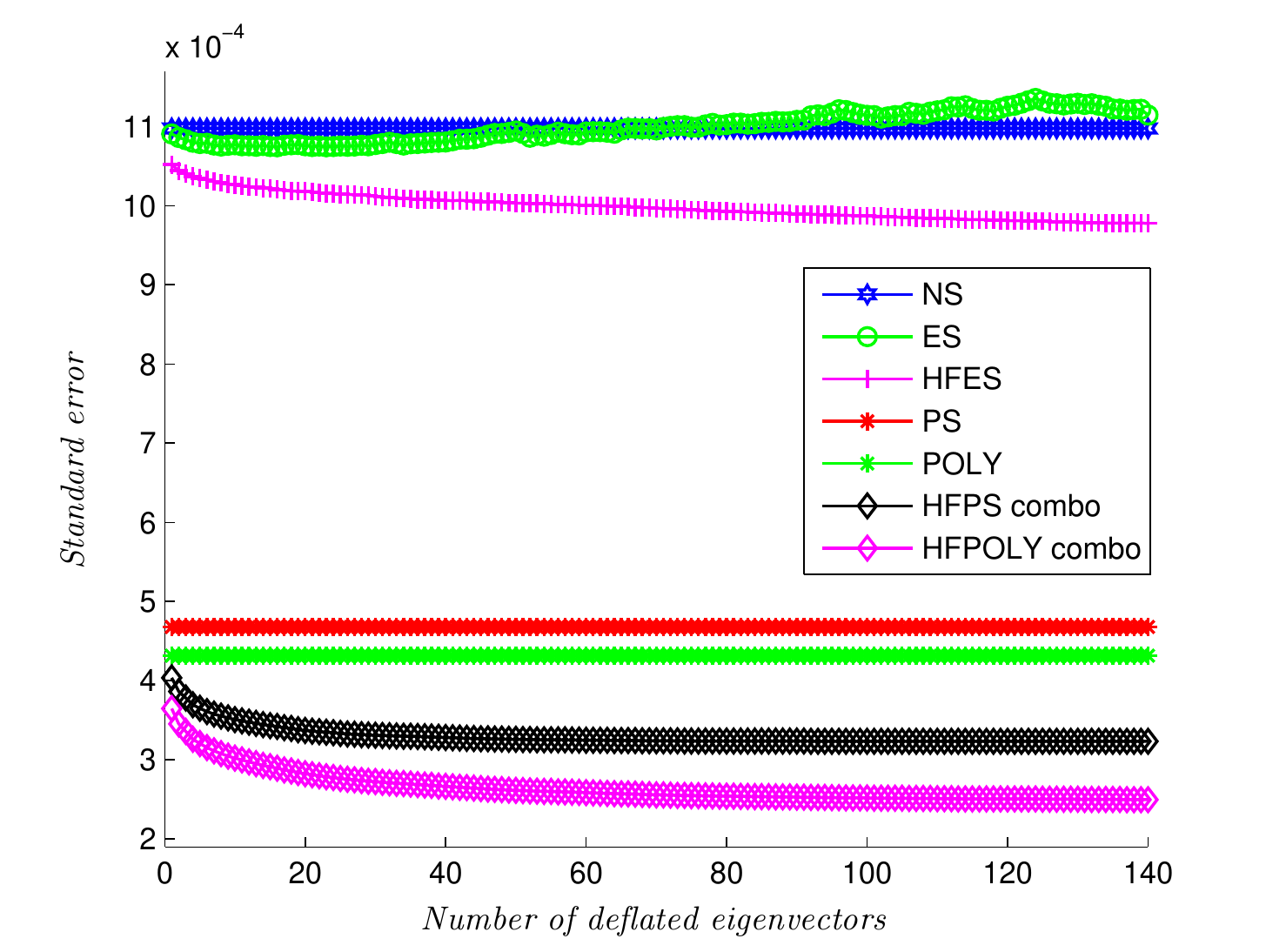}
\caption{Quenched $8^4$ lattice simulation results for the standard error as a function of deflated eigenvalues for the $J_4$ point-split operator at $\kappa = 0.157$.}
\label{figns74}
\end{figure}
\begin{figure}[!htpb]
\centering
\includegraphics[trim={1.1cm 8cm 0 8.5cm},clip,width=1.10\textwidth]{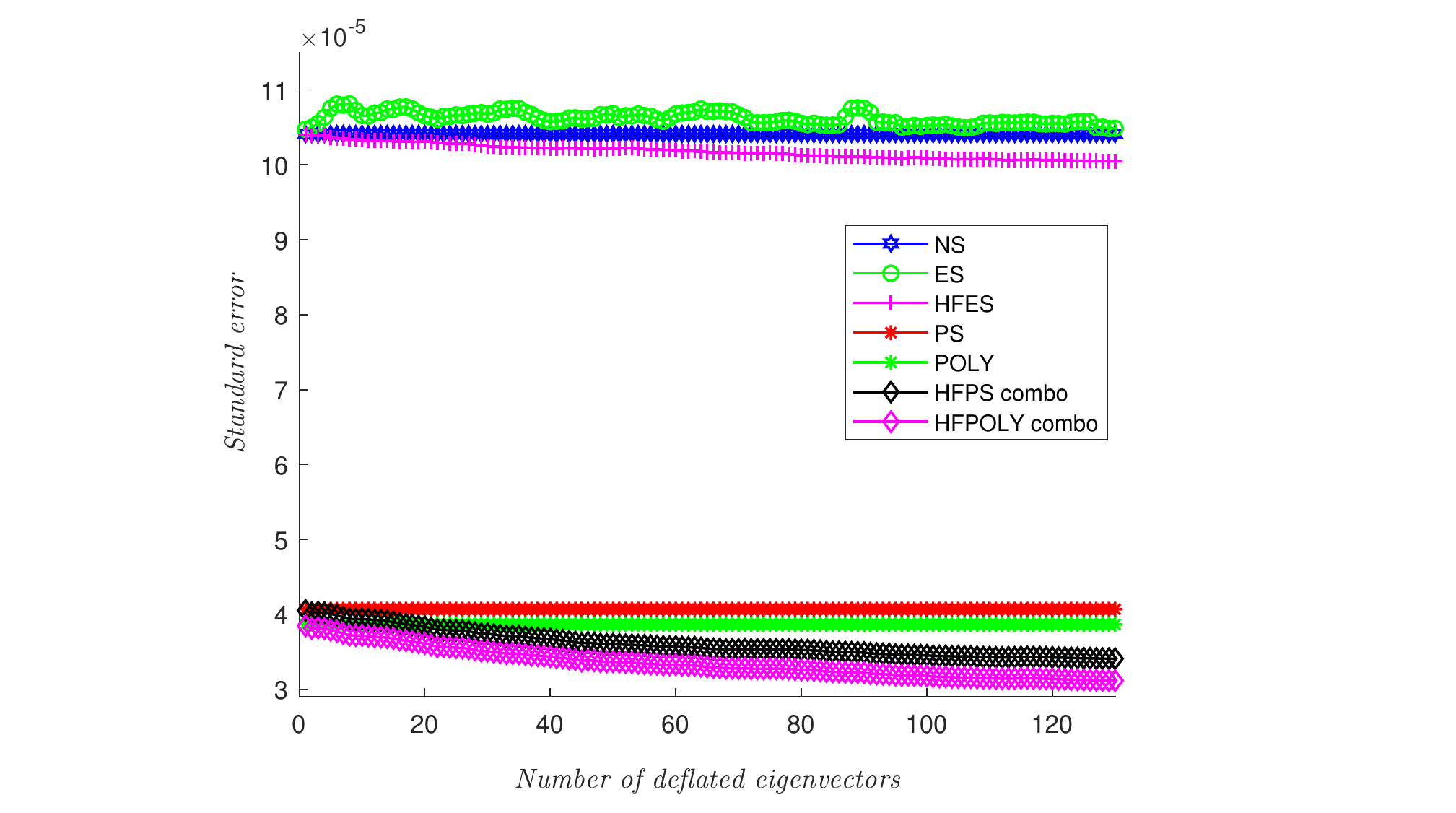}
\vspace{-.75cm}
\caption{Quenched $24^3\times32$ lattice simulation results for the standard error as a function of deflated eigenvalues for the $J_4$ point-split operator at $\kappa = 0.155$.}
\label{fignl54}
\end{figure}
\begin{figure}[!htpb]
\centering
\includegraphics[trim={1.1cm 8cm 0 8.5cm},clip,width=1.10\textwidth]{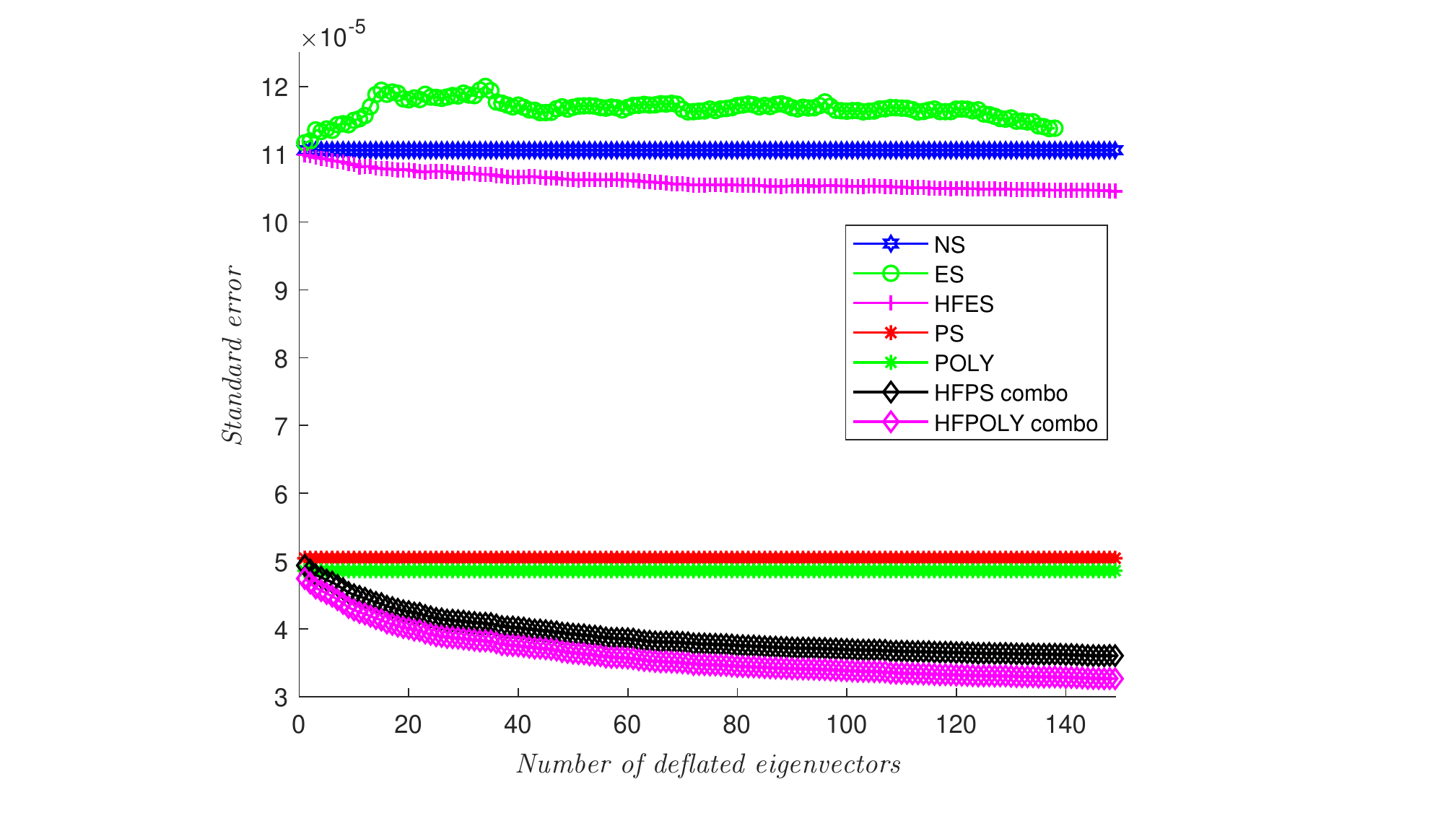}
\vspace{-.75cm}
\caption{Quenched $24^3\times32$ lattice simulation results for the standard error as a function of deflated eigenvalues for the $J_4$ point-split operator at $\kappa = 0.156$.}
\label{fignl64}
\end{figure}

Figs.~\ref{figns54}, \ref{figns64} and  \ref{figns74} display the corresponding error bars for the point-split vector calculated with $\kappa = 0.155$, 0.156 and 0.157, respectively, for the $8^4$ lattice. Similarly, Figs \ref{fignl54} and \ref{fignl64} display the calculated error bars for the point-split vector with $\kappa = 0.155$ and 0.156 for the $24^3 \times32$ lattice. The qualitative behavior of the statistical error bars for the point-split vector is similar to that of the local vector, however, there is a larger separation between the PS and POLY methods, leading to an increased separation between HFPS and HFPOLY methods than for the local operator case. 
We again observe the trend that the separation between the calculated error bars of the HFPS/HFPOLY methods and the PS method increases as $\kappa$ is increased.

\begin{figure}[!htpb]
\centering
\includegraphics[width=.75\textwidth]{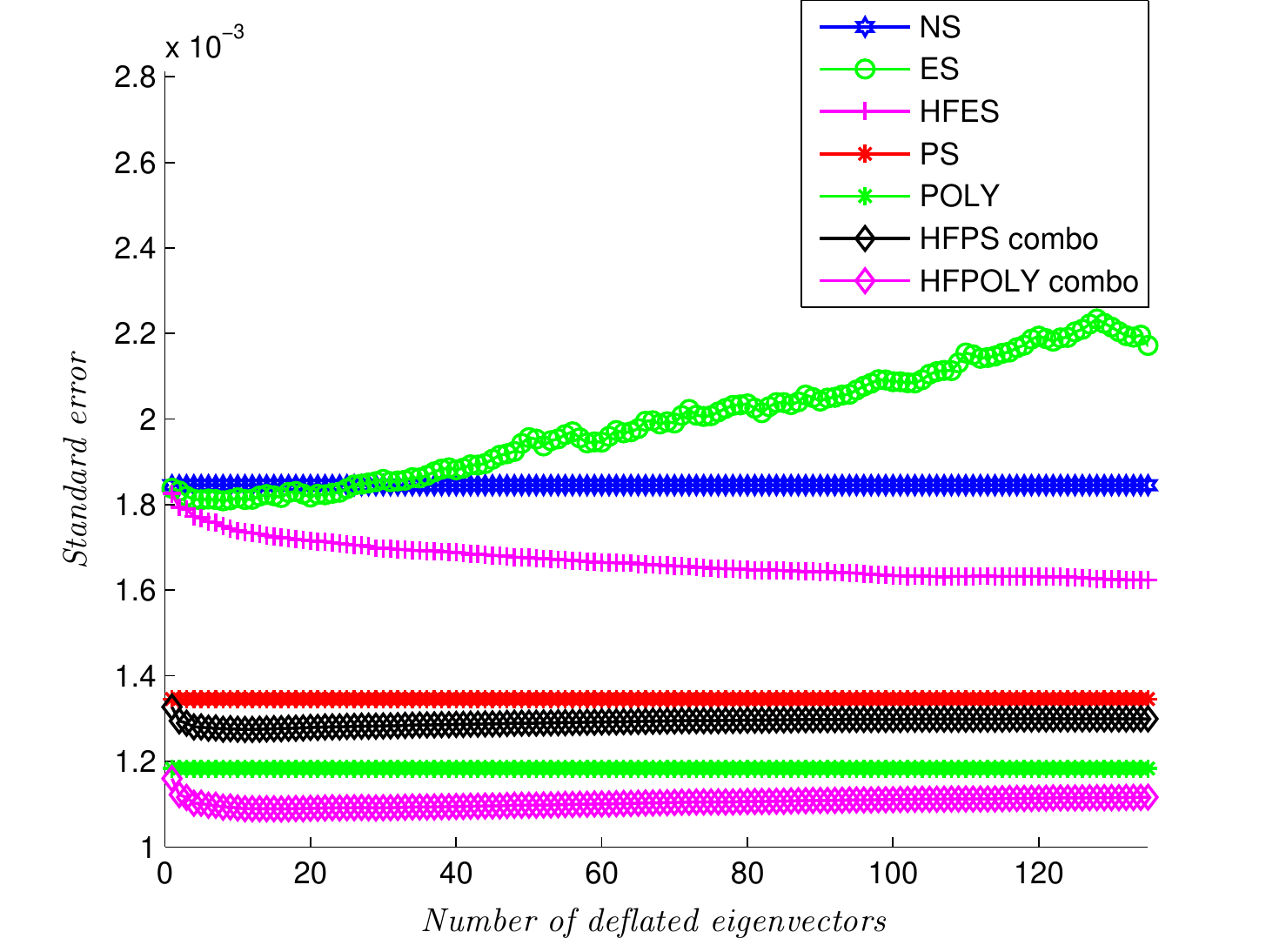}
\caption{Quenched $8^4$ lattice simulation results for the standard error as a function of deflated eigenvalues for the scalar operator at $\kappa = 0.155$.}
\label{figs55}
\end{figure}

\begin{figure}[!htpb]
\centering
\includegraphics[width=.75\textwidth]{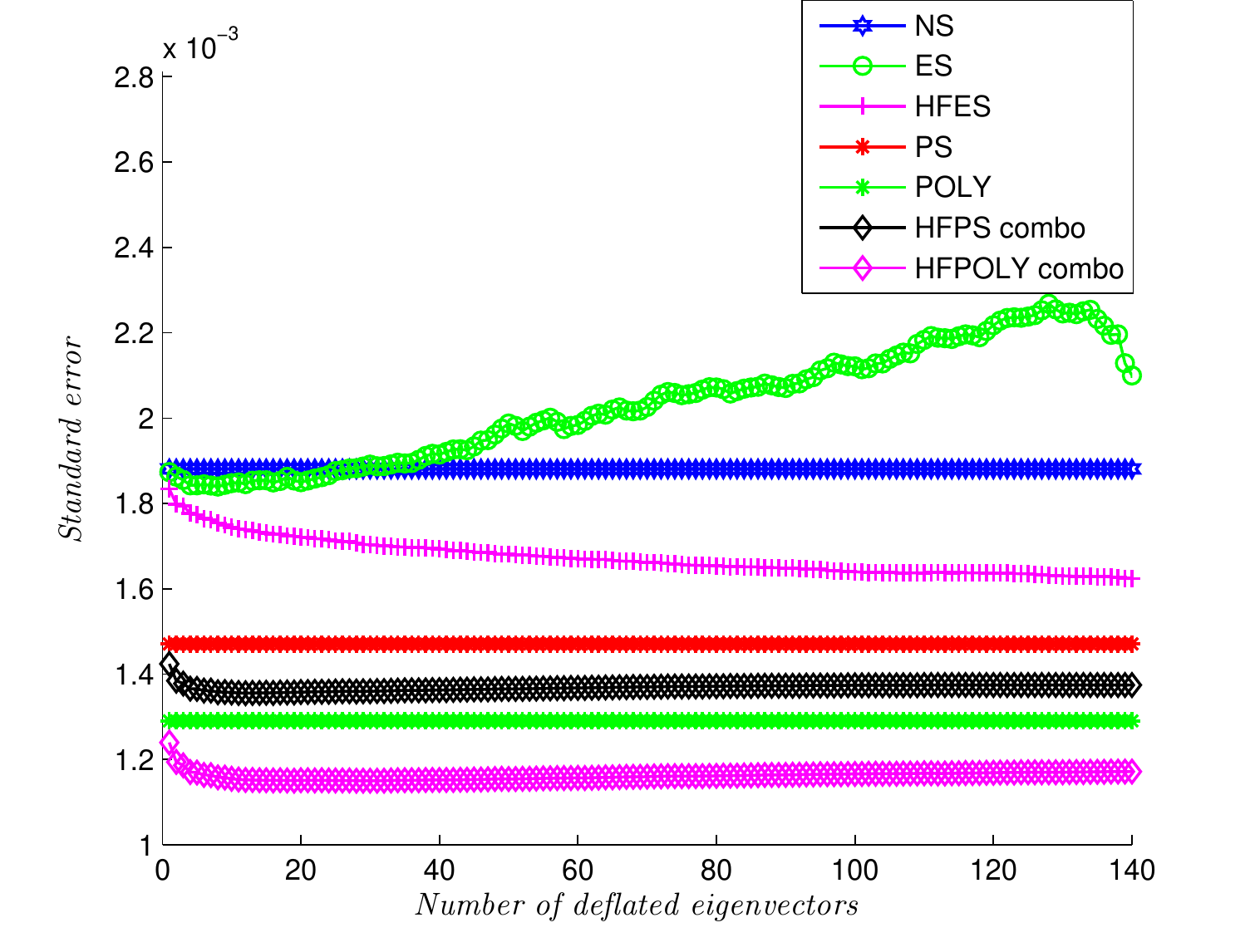}
\caption{Quenched $8^4$ lattice simulation results for the standard error as a function of deflated eigenvalues for the scalar operator at $\kappa = 0.156$.}
\label{figs65}
\end{figure}

\begin{figure}[!htpb]
\centering
\includegraphics[width=.75\textwidth]{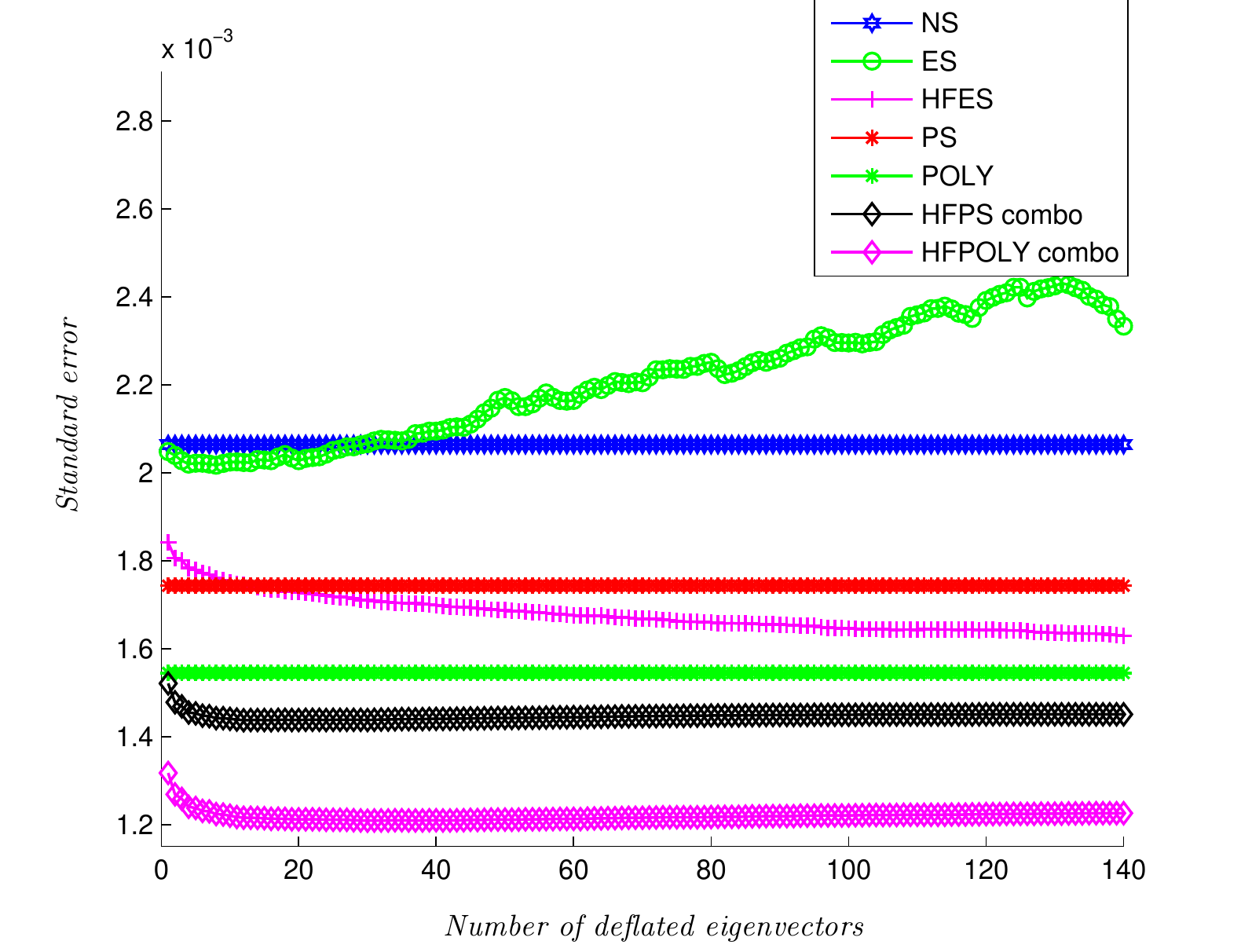}
\caption{Quenched $8^4$ lattice simulation results for the standard error as a function of deflated eigenvalues for the scalar operator at $\kappa = 0.157$.}
\label{figs75}
\end{figure}

\begin{figure}[!htpb]
\centering
\includegraphics[trim={1.1cm 8cm 0 8.5cm},clip,width=1.10\textwidth]{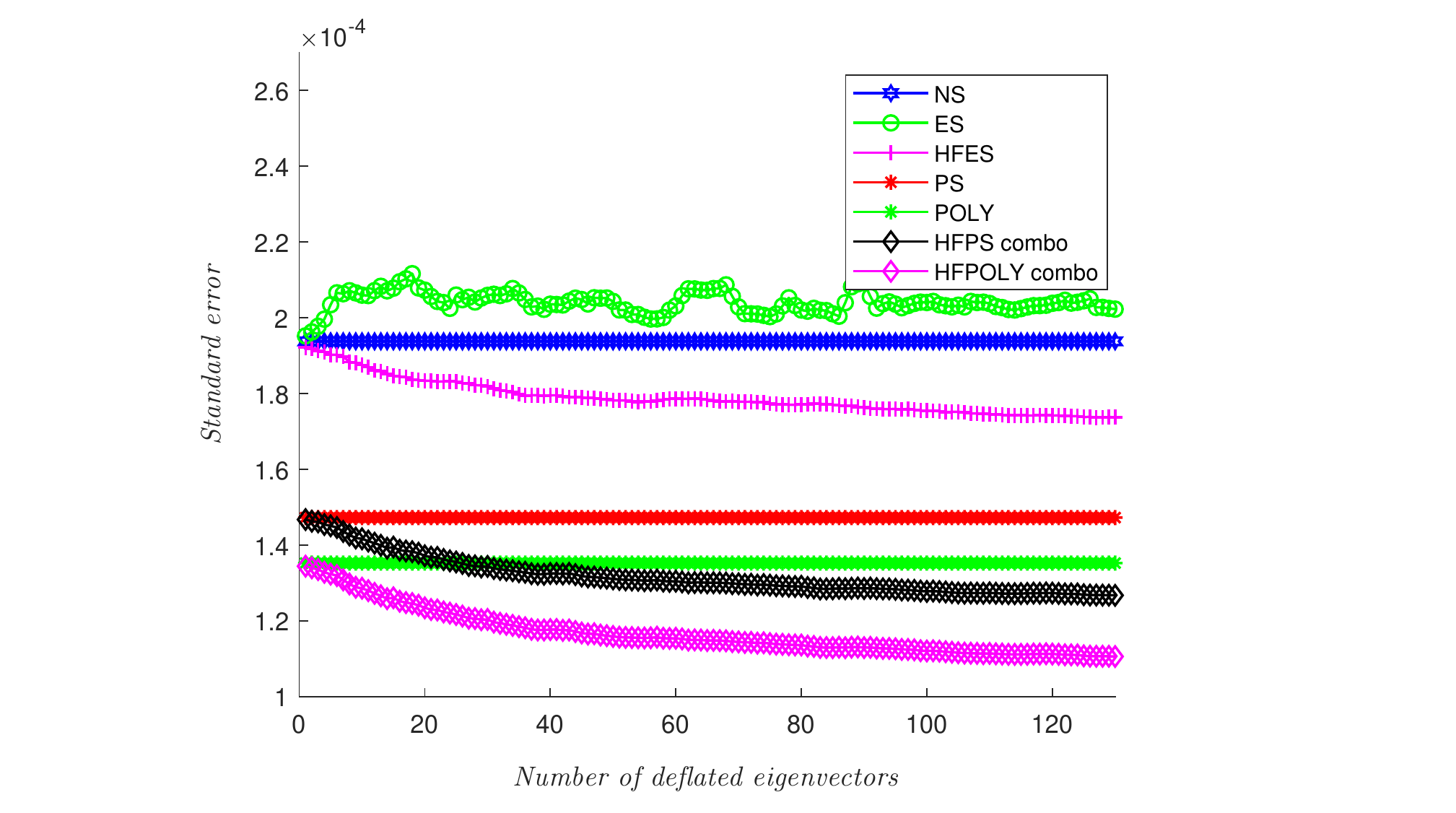}
\vspace{-.75cm}
\caption{Quenched $24^3\times32$ lattice simulation results for the standard error as a function of deflated eigenvalues for the scalar operator at $\kappa = 0.155$.}
\label{figl55}
\end{figure}

\begin{figure}[!htpb]
\centering
\includegraphics[trim={1.1cm 8cm 0 8.5cm},clip,width=1.10\textwidth]{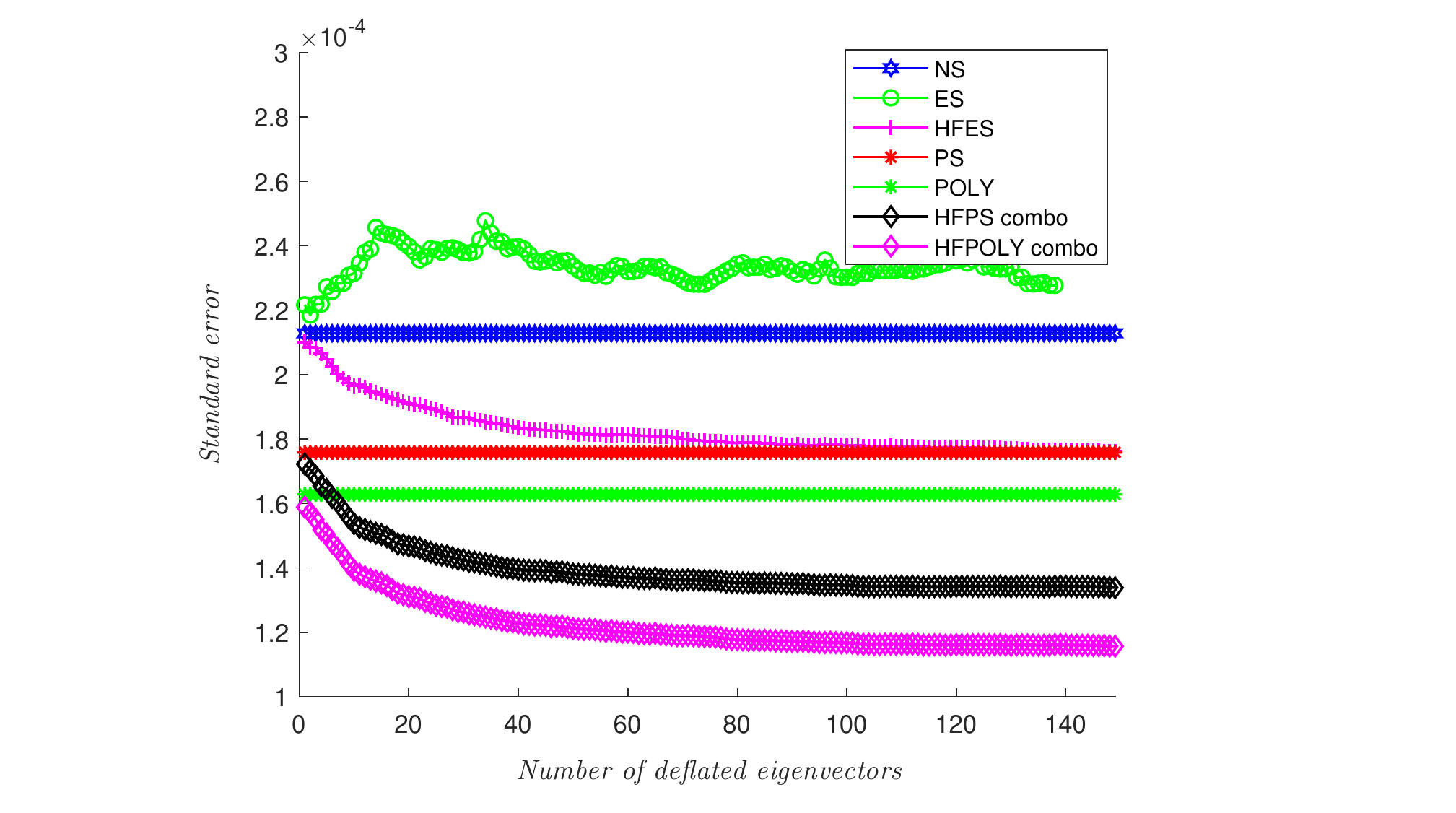}
\vspace{-.75cm}
\caption{Quenched $24^3\times32$ lattice simulation results for the standard error as a function of deflated eigenvalues for the scalar operator at $\kappa = 0.156$.}
\label{figl65}
\end{figure}

Figs.~\ref{figs55}, \ref{figs65} and \ref{figs75} display the statisical error bars calculated for the scalar operator at $\kappa = 0.155$, 0.156 and 0.157, respectively, for the $8^4$ lattice. Similarly, Figs.~\ref{figl55} and \ref{figl65} display the statistical error bars for the scalar operator on the $24^3 \times32$ lattice, at $\kappa = 0.155$ and 0.156, respectively. The separation between the error bars of the POLY and PS methods has increased for the scalar operator, resulting in an increase of the separation between the HFPS and HFPOLY methods for both lattice volumes and all values of $\kappa$. We also observe the general trend that the separation between the HFPS/HFPOLY methods and the PS method increases as $\kappa$ increases in value for the scalar operator. The HFES method produces error bars that decrease below those produced through the PS method on the $8^4$ lattice at $\kappa=0.157$, as can be seen from Fig.~\ref{figs75}. A discontinuity is also observed for all the the hermitian methods at both $\kappa=0.156, 0.157$ on the $8^4$ lattice for the scalar operator, as was also seen for the local and point-split vector operators. 

We define the relative efficiency, $RE$, of two methods as
\begin{equation}
RE\equiv \left( \frac {1}{\delta y^{2}}-1\right) \times 100,
\end{equation}
where $\delta y$ is the relative error bar ratio and $\delta y^2$ is the relative variance. Since the variance is proportional to the square of the statistical error, $RE$ actually compares relative variance efficiencies. Variances are the significant variable to consider since computer time for the noise calculation is controlled by the operator variance. Of course an increase in the relative efficiency between two methods implies a decrease in the variance of the tested method. Tables \ref{1550small}-\ref{1560large} show the relative efficiency as calculated for both lattice volumes and all values of $\kappa$ for each of the three operators. Here, we compare the POLY, HFES, HFPS and HFPOLY methods to the NS and PS methods. The HFPS and HFPOLY methods outperform the PS method across all lattice volumes and all values of $\kappa$, with the HFPOLY method increasing the relative efficiency the most out of all subtraction methods.  A comparison of relative efficiency across all operators shows that the relative efficiency of HFPOLY {\em vs.} PS increases as the quark mass decreases towards physical masses. This is to be expected as the perturbative approximation to the inverse of the Wilson matrix becomes less accurate at larger values of $\kappa$. We also observe a deflation saturation effect on our hermitian subtraction methods similar to the slope saturation seen in Ref.~\cite{Wilcox2} for quark propagators, wherein the slope of noise errors {\em vs.} subtracted eigenvalues quickly flatens out. Due to this phenomenon, only a small number of eigenmodes need be calculated to achieve a significant reduction in the variance, without the increased cost of additional matrix vector products.

Note that there is a consistent pattern in the effectiveness of HFES across operators and lattices in these results. As seen in all the tables ({\lq\lq\em vs.}\,\,NS"), the scalar and local vector operators respond similarly to deflation, whereas the point-split vector is always less responsive. Also seen in these tables (\lq\lq{\em vs.} PS") are the situations where subtraction from HFES begins to beat or is essentially the same as PS: the scalar operator at $\kappa=0.157$ on the $8^4$ lattices (Fig.~\ref{figs75}), the scalar operator at $\kappa=0.156$ on the $24^3\times 32$ lattices (Fig.~\ref{figl65}), all the operators at $\kappa_{crit}$ on the $24^3\times 32$ lattices, and all the dynamical results, except the point-split vector, which again responds less well to deflation subtraction. Clearly, deflation effectiveness is increasing as quark mass is decreased. We will see a strong confirmation of this trend in the next section.


\begin{table}[H]
\centering
\begin{tabular}{ c|c|c|c|c|c|c| } 
\cline{2-7}
\multicolumn{1}{c|}{}& \multicolumn{2}{c|}{Scalar}&\multicolumn{2}{c|}{Local $J_4$}&\multicolumn{2}{c|}{Point-Split $J_4$}\\
\hline
\multicolumn{1}{|c|}{Subtraction} & {\em{vs.}} NS  & {\em{vs.}} PS  &  {\em{vs.}} NS  & {\em{vs.}} PS &  {\em{vs.}} NS  &  {\em{vs.}} PS\\ 
\hline
\multicolumn{1}{|c|}{POLY} & $143\%$ &  $29.4\%$   & $274\%$ &  $1.8\%$ & $778\%$ &  $15.7\%$ \\
\hline 
\multicolumn{1}{|c|}{HFES} & $29.9\%$ &  $-30.9\%$    & $48.7\%$ &  $-59.5\%$ & $16.6\%$ &  $-84.7\%$ \\
\hline
\multicolumn{1}{|c|}{HFPS} & $102\%$ &  $7.3\%$    & $978\%$ &  $194\%$ & $1230\%$ &  $74.7\%$ \\
\hline
\multicolumn{1}{|c|}{HFPOLY} & $173\%$ &  $45.3\%$    & $1620\%$ &  $369\%$ & $2030\%$ &  $181\%$ \\
\hline
\end{tabular}
 \caption{Comparison of relative efficiency for operators on the $8^4$ lattice at $\kappa$ = 0.155.}
\label{1550small}
\vspace{3cm}
\begin{tabular}{ c|c|c|c|c|c|c| } 
\cline{2-7}
\multicolumn{1}{c|}{}& \multicolumn{2}{c|}{Scalar}&\multicolumn{2}{c|}{Local $J_4$}&\multicolumn{2}{c|}{Point-Split $J_4$}\\
\hline
\multicolumn{1}{|c|}{Subtraction} & {\em{vs.}} NS  & {\em{vs.}} PS  &  {\em{vs.}} NS  & {\em{vs.}} PS &  {\em{vs.}} NS  &  {\em{vs.}} PS\\   
\hline
\multicolumn{1}{|c|}{POLY}  & $112\%$ &  $29.8\%$  & $245\%$ &  $2.8\%$ & $678\%$ &  $17.1\%$  \\
\hline 
\multicolumn{1}{|c|}{HFES} & $34.1\%$ &  $-18.0\%$   & $53.7\%$ &  $-54.2\%$ & $18.0\%$ &  $-82.2\%$  \\
\hline
\multicolumn{1}{|c|}{HFPS} & $87.1\%$ &  $14.4\%$  & $937\%$ &  $209\%$ & $1100\%$ &  $80.8\%$ \\
\hline
\multicolumn{1}{|c|}{HFPOLY} & $158\%$ &  $57.5\%$  & $1610\%$ &  $409\%$  & $1880\%$ &  $197\%$ \\
\hline
\end{tabular}
\caption{Comparison of relative efficiency for operators on the $8^4$ lattice at $\kappa$ = 0.156.}
\label{1560small}
\vspace{3cm}
\begin{tabular}{ c|c|c|c|c|c|c| } 
\cline{2-7}
\multicolumn{1}{c|}{}& \multicolumn{2}{c|}{Scalar}&\multicolumn{2}{c|}{Local $J_4$}&\multicolumn{2}{c|}{Point-Split $J_4$}\\
\hline
\multicolumn{1}{|c|}{Subtraction} & {\em{vs.}} NS  & {\em{vs.}} PS  &  {\em{vs.}} NS  & {\em{vs.}} PS &  {\em{vs.}} NS  &  {\em{vs.}} PS\\   
\hline
\multicolumn{1}{|c|}{POLY}  & $78.6\%$ &  $27.4\%$  & $202\%$ &  $3.6\%$ & $547\%$ &  $17.6\%$ \\
\hline 
\multicolumn{1}{|c|}{HFES}  & $60.5\%$ &  $14.5\%$   & $74.2\%$ &  $-40.1\%$ & $26.0\%$ &  $-77.1\%$ \\
\hline
\multicolumn{1}{|c|}{HFPS}   & $103\%$ &  $44.5\%$ & $993\%$ &  $276\%$ & $1050\%$ &  $109\%$ \\
\hline
\multicolumn{1}{|c|}{HFPOLY}   & $184\%$ &  $102\%$  & $1760\%$ &  $538\%$ & $1830\%$ &  $252\%$ \\
\hline
\end{tabular}
\caption{Comparison of relative efficiency for operators on the $8^4$ lattice at $\kappa$ = 0.157.}
\label{1570small}
\end{table}
\vspace{2cm}
\begin{table}[h!]
\centering
\begin{tabular}{ c|c|c|c|c|c|c|} 
\cline{2-7}
\multicolumn{1}{c|}{}& \multicolumn{2}{c|}{Scalar}&\multicolumn{2}{c|}{Local $J_4$}&\multicolumn{2}{c|}{Point-Split $J_4$}\\
\hline
\multicolumn{1}{|c|}{Subtraction} & {\em{vs.}} NS  & {\em{vs.}} PS  &  {\em{vs.}} NS  & {\em{vs.}} PS &  {\em{vs.}} NS  &  {\em{vs.}} PS\\ 
\hline
\multicolumn{1}{|c|}{POLY}  & $105\%$ &  $18.5\%$   & $194\%$ &  $-0.53\%$ & $625\%$ &  $10.7\%$ \\
\hline 
\multicolumn{1}{|c|}{HFES}  & $24.4\%$ &  $-28.2\%$   & $21.1\%$ &  $-59.0\%$ & $7.4\%$ &  $-83.6\%$ \\
\hline
\multicolumn{1}{|c|}{HFPS}  & $134\%$ &  $35.0\%$ & $409\%$ &  $72.3\%$ & $833\%$ &  $42.5\%$ \\
\hline
\multicolumn{1}{|c|}{HFPOLY} & $207\%$ &  $77.2\%$  & $421\%$ &  $76.1\%$ & $1020\%$ &  $70.5\%$ \\
\hline
\end{tabular}
\caption{Comparison of relative efficiency for operators on the $24^3\times 32$ lattice at $\kappa$ = 0.155.}
\label{1550large}
\end{table}
\begin{table}[h!]
\centering
\begin{tabular}{ c|c|c|c|c|c|c| } 
\cline{2-7}
\multicolumn{1}{c|}{}& \multicolumn{2}{c|}{Scalar}&\multicolumn{2}{c|}{Local $J_4$}&\multicolumn{2}{c|}{Point-Split $J_4$}\\
\hline
\multicolumn{1}{|c|}{Subtraction} & {\em{vs.}} NS  & {\em{vs.}} PS  &  {\em{vs.}} NS  & {\em{vs.}} PS &  {\em{vs.}} NS  &  {\em{vs.}} PS\\  
\hline
\multicolumn{1}{|c|}{POLY} & $70.8\%$ &  $16.6\%$   & $129\%$ &  $1.3\%$ & $418\%$ &  $7.7\%$ \\
\hline 
\multicolumn{1}{|c|}{HFES}& $45.9\%$ &  $-0.39\%$  & $35.7\%$ &  $-39.9\%$ & $11.9\%$ &  $-76.7\%$  \\
\hline
\multicolumn{1}{|c|}{HFPS} & $153\%$ &  $72.6\%$  & $460\%$ &  $148\%$ & $841\%$ &  $95.6\%$ \\
\hline
\multicolumn{1}{|c|}{HFPOLY} & $238\%$ &  $131\%$  & $479\%$ &  $156\%$ & $1050\%$ &  $138\%$ \\
\hline
\end{tabular}
 \caption{Comparison of relative efficiency for operators on the $24^3\times 32$ lattice at $\kappa$ = 0.156.}
\label{1560large}
\end{table}

\section{Subtraction Methods at $\kappa_{crit}$}\label{Seight}

\begin{figure}[!htpb]
\centering
\includegraphics[trim={4cm 2.5cm 0cm 2cm},clip,width=1.25\textwidth]{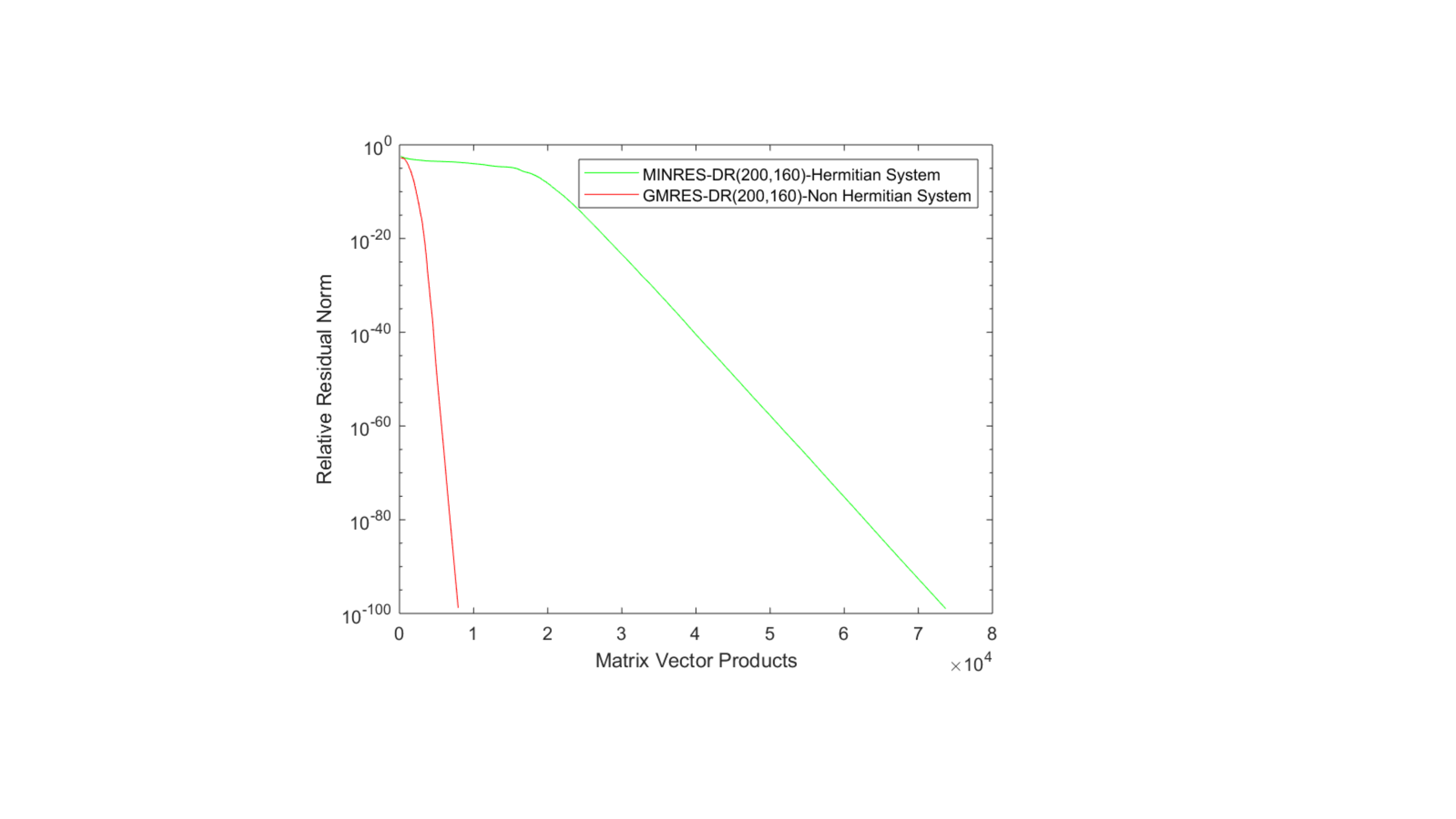}
\vspace{-.75cm}
\caption{The performance of the MINRES and GMRES algorithms at $\kappa_{crit}$ on the hermitian and nonhermitian systems.}
\label{GMRES_MINRES}
\end{figure}

Due to the difficulty of solving the linear and eigenvalue equations of the large hermitian Wilson matrix at $\kappa_ {crit}$, MINRES-DR was used to calculate the eigenmodes used for subtraction. MINRES-DR is an iterative solver specific for hermitian, indefinite systems, and as such is perfectly suited for these methods at $\kappa_{crit}$ \cite{Nicely,MINRESDR}. At smaller lattice volumes and smaller values of $\kappa$, GMRES-DR and MINRES-DR perform identically in solving the linear equations and eigenvalue equations of the hermitian matrix. However, GMRES-DR suffers from numerical instability that hampers convergence at $\kappa_{crit}$, while MINRES-DR does not. The approximate value of $\kappa_{crit}$ for quenched Wilson lattices at $\beta=6.0$ is given in Ref.~\cite{Cabasino} as 0.15708(2). We adopted $\kappa_{crit}=0.157$ for this study. The linear equations were computed to super-convergence to ensure that a sufficient number of eigenmodes were obtained for this study. 130 eigenmodes of the hermitian Wilson matrix were obtained across all ten configurations, and were used in our subtraction methods. Fig.~\ref{GMRES_MINRES} shows the super-convergence performance of the residual norm of the GMRES-DR and MINRES-DR algorithms as a function of matrix vector products at $\kappa=0.157$. Both curves show the onset of deflation as described in Ref.~\cite{Wilcox2}, i.e., a pronounced \lq\lq knee" followed by a smooth linear semi-log decrease. This is characteristic of our calculations at all $\kappa$ values, although the number of matrix-vector products increases sharply as $\kappa$ increases toward $\kappa_{crit}$. It is important use use deflated linear equations solvers in any lattice calculation at low quark mass.

\begin{figure}[!htpb]
\centering
\includegraphics[trim={1.1cm 8cm 0 8.5cm},clip,width=1.10\textwidth]{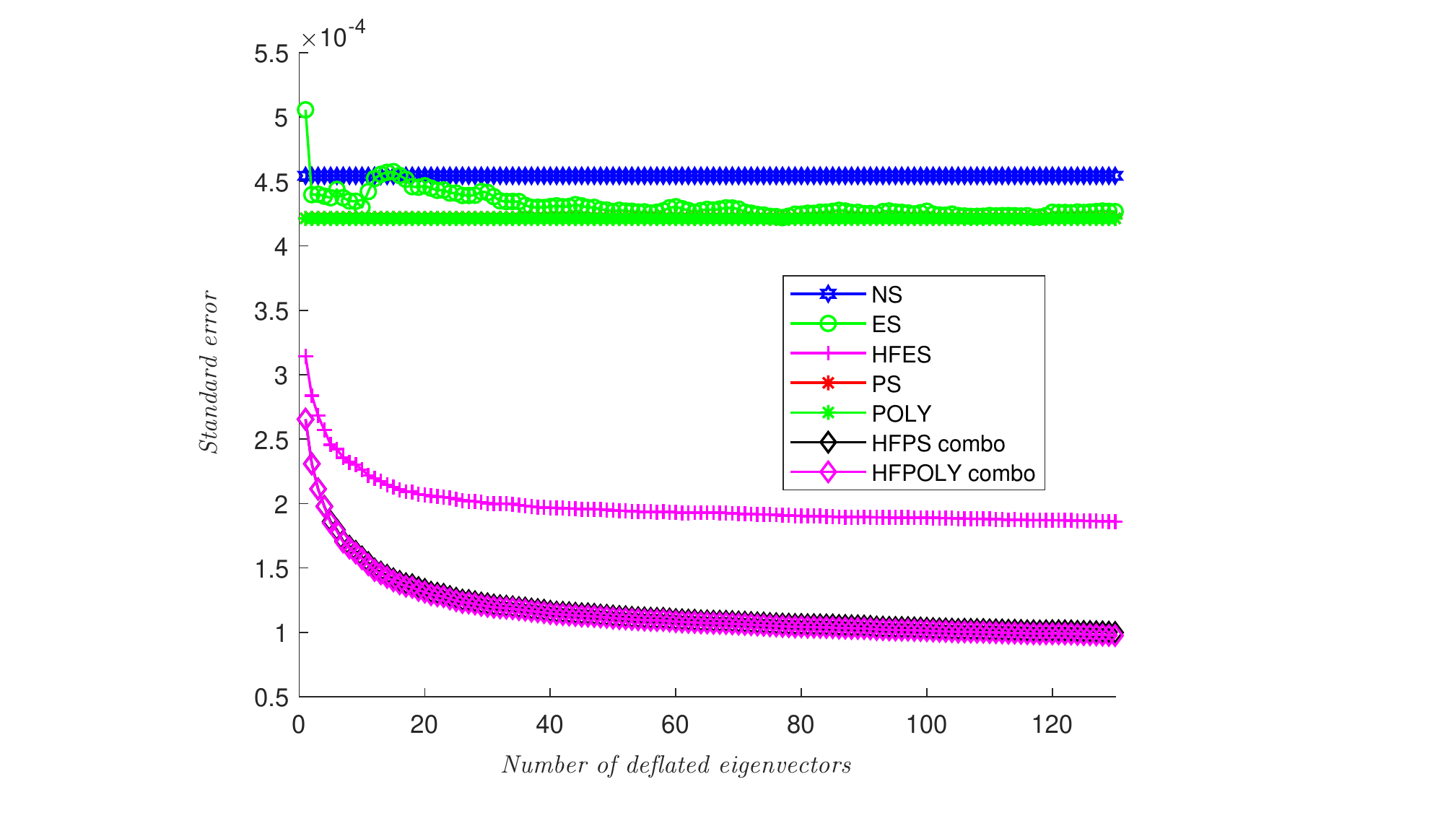}
\vspace{-.75cm}
\caption{Quenched $24^3\times32$ lattice simulations for the standard error as a function of deflated eigenvalues for the $J_4$ local operator at $\kappa_{crit}$.}
\label{Tigs2}
\end{figure}

\begin{figure}[!htpb]
\centering
\includegraphics[trim={1.1cm 8cm 0 8.5cm},clip,width=1.10\textwidth]{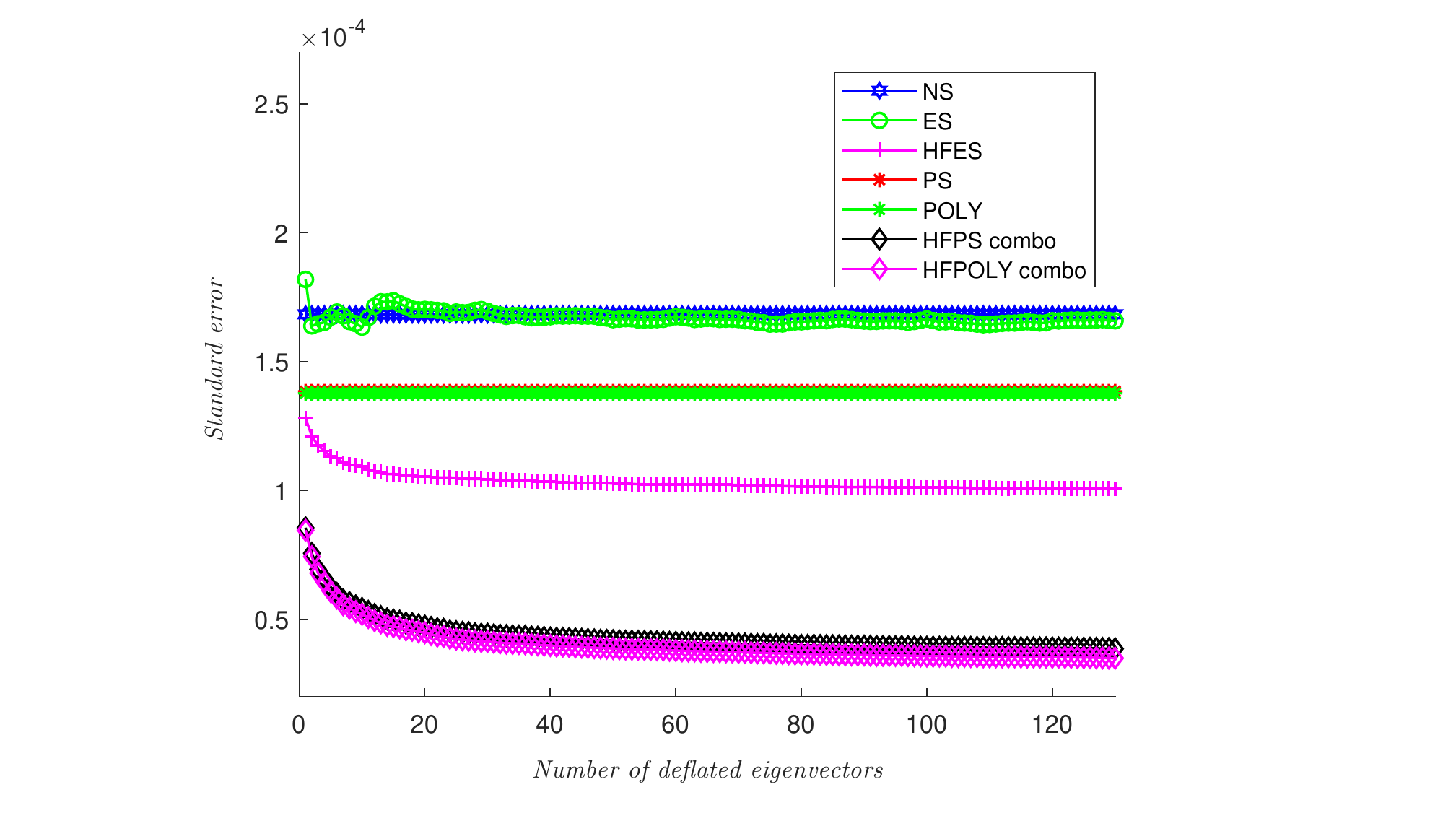}
\vspace{-.75cm}
\caption{Quenched $24^3\times32$ lattice simulations for the standard error as a function of deflated eigenvalues for the $J_4$ point-split operator at $\kappa_{crit}$.}
\label{Tigs1}
\end{figure}


\begin{figure}[!htpb]
\centering
\includegraphics[trim={1.1cm 8cm 0 8.5cm},clip,width=1.10\textwidth]{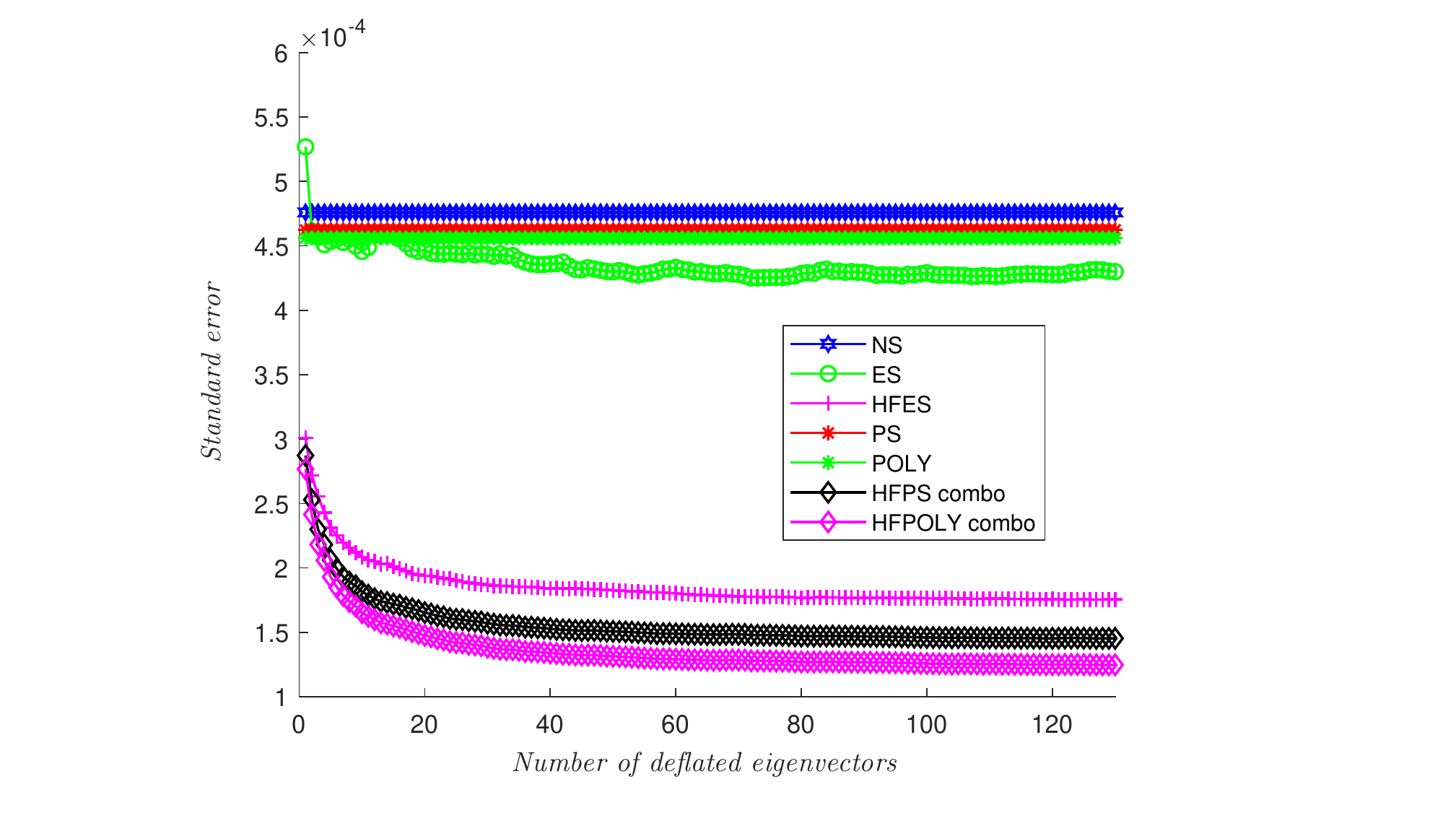}
\vspace{-.75cm}
\caption{Quenched $24^3\times32$ lattice simulations for the standard error as a function of deflated eigenvalues for the scalar operator at $\kappa_{crit}$.}
\label{Tigs3}
\end{figure}

Figs.~\ref{Tigs2}, \ref{Tigs1} and \ref{Tigs3} show the error bars as a function of the number of deflated eigenvalues for the local, point-split and scalar operators, respectively. We see that at low quark mass subtraction methods that deflate out the hermitian eigenvalues become much more effective than the standard method of perturbative subtraction. The perturbative and polynomial approximations to the inverse of the Wilson matrix become less accurate at $\kappa_{crit}$, and the subtraction methods using these approximations become less efficient in this limit. In the case of the scalar and local charge density operators, the error bars of these methods approach those of the NS method.

At $\kappa_{crit}$, we observe strong low eigenvalue dominance as the HF error bars dramatically decrease for all operators. Interestingly, the HFES method becomes much more effective than either PS or POLY for the scalar operator. Also notice the following enhancement effect evident in Figs.~\ref{Tigs2}, \ref{Tigs1} and \ref{Tigs3}: the HFPS and HFPOLY {\em vs.} HF effects are enhanced relative to PS and POLY {\em vs.} NS. That is, when deflation is included, PS and POLY methods are more effective. This enhancement effect is similar to that seen in Ref.~\cite{Gambhir} for HP combined with deflation {\em vs.} HP alone.

Of importance to note is the role of the lowest lying eigenmodes of the hermitian matrix. As can be seen for simulations on the lattice of size $8^4$, a discontinuity is observed for the hermitian subtractions methods at $\kappa_{crit}$ (see Figs.~\ref{figs74}, \ref{figns74} and \ref{figs75}). The same discontinuity is observed on lattices of a larger volume at $\kappa_{crit}$, however, it is much more pronounced. A significant decrease in the value of the error bars is observed upon subtraction of the lowest lying eigenvalue in the case of HFES, HFPS, and HFPOLY methods in comparison to the NS, PS and POLY methods, respectively. Again note that computing a large number of eigenmodes is not necessary for the hermitian methods to be effective. The error bars experience deflation saturation after $\sim$ 20-30 eigenmodes, decreasing the number of matrix vector products required from a super-convergence step with little loss of effectiveness.

\begin{table}
\centering
\begin{tabular}{ c|c|c|c|c|c|c| } 
\cline{2-7}
\multicolumn{1}{c|}{}& \multicolumn{2}{c|}{Scalar}&\multicolumn{2}{c|}{Local $J_4$}&\multicolumn{2}{c|}{Point-Split $J_4$}\\
\hline
\multicolumn{1}{|c|}{Subtraction} & {\em{vs.}} NS  & {\em{vs.}} PS  &  {\em{vs.}} NS  & {\em{vs.}} PS &  {\em{vs.}} NS  &  {\em{vs.}} PS\\  
\hline
\multicolumn{1}{|c|}{POLY} & $8.9\%$ & $2.8\%$ & $16.4\%$ & $0.1\%$ & $49.5\%$ & $1.1\%$  \\
\hline 
\multicolumn{1}{|c|}{HFES} & $634\%$ & $593\%$ & $496\%$ & $413\%$ & $180\%$ & $89.2\%$ \\
\hline
\multicolumn{1}{|c|}{HFPS} & $972\%$ & $911\%$ & $1970\%$ & $1680\%$ & $1800\%$ & $1180\%$  \\
\hline
\multicolumn{1}{|c|}{HFPOLY} & $1350\%$ & $1270\%$ & $2070\%$ & $1770\%$  & $2220\%$ & $1470\%$  \\
\hline
\end{tabular}
 \caption{Comparison of relative efficiencies for operators at $\kappa_{crit}$ on the quenched lattice of size $24^3\times32$.}
\label{table1}
\end{table}

Table \ref{table1} displays the relative efficiencies of the POLY, HFES, HFPS and HFPOLY subtraction methods in comparison to the NS and PS methods for the scalar, local and point-split operators at $\kappa_{crit}$. As can be seen for all three operators, the HFES, HFPS and HFPOLY methods are much more effective than both the NS and PS methods. 
The relative efficiency between the HFPOLY and PS methods is over $1200\%$ for all three operators. This implies a decrease in the relative variance, $\delta y^2$, by over an order of magnitude. Thus, using the HFPOLY method will decrease the computer time for simulations by the same amount when using noise methods.


\section{Dynamical Lattices}\label{Snine}

\begin{figure}[t!]
\centering
\includegraphics[width=.65\textwidth]{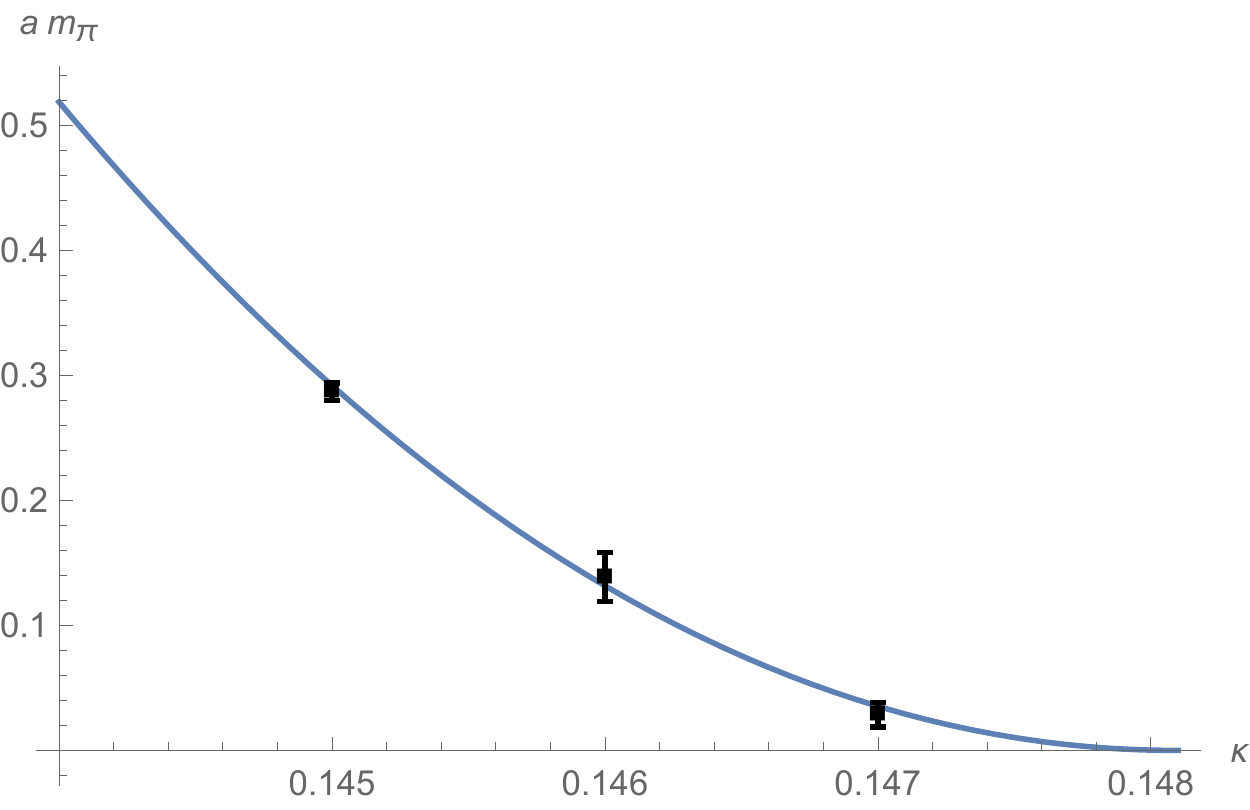}
\caption{The mass of the pion in dimensionless units as a function of the hopping parameter.}
\label{figs4}
\includegraphics[trim={1cm 8cm 0 8cm},clip,width=1.10\textwidth]{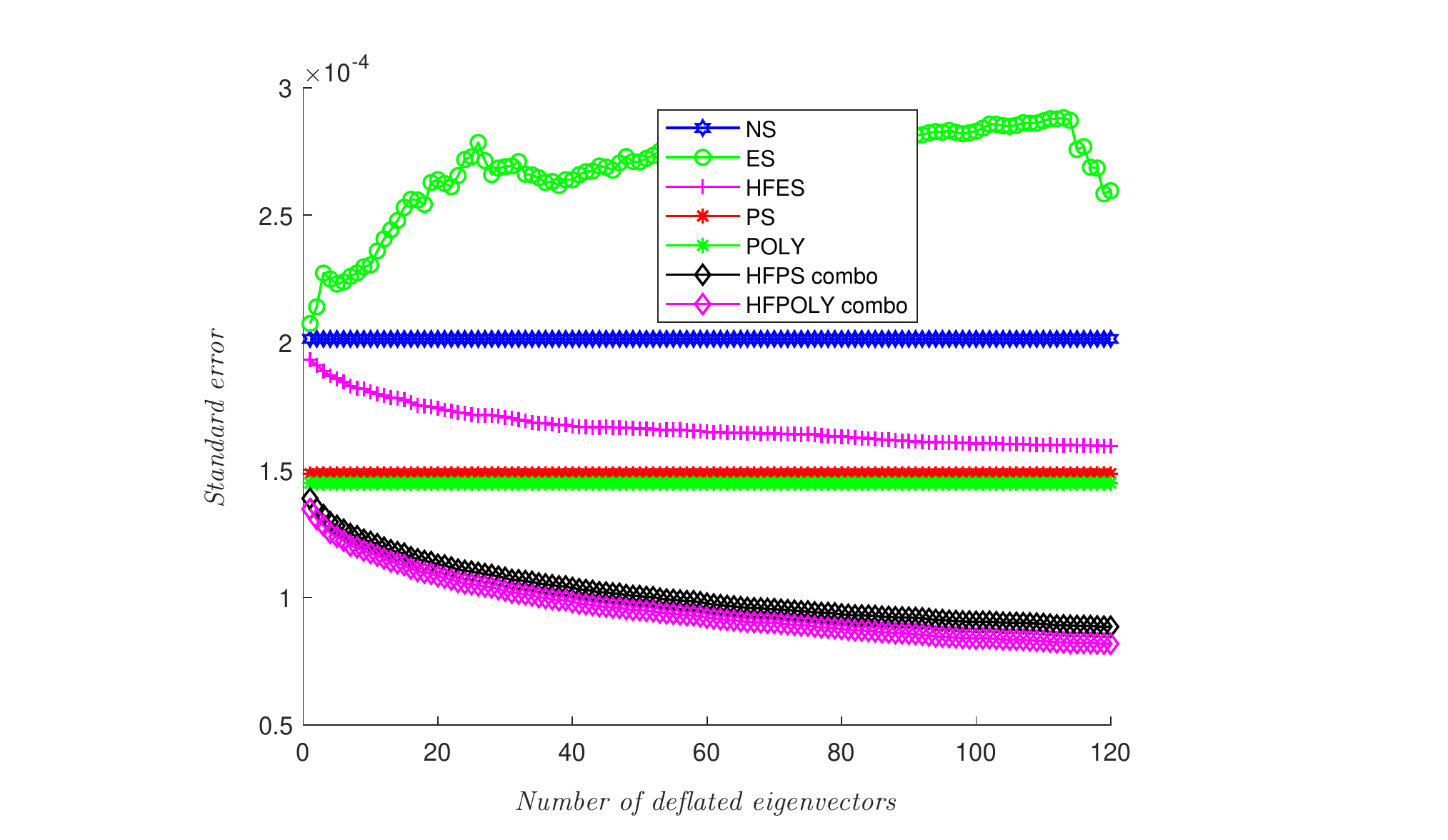}
\vspace{-.75cm}
\caption{Dynamical $16^3\times48$ lattice simulations for the standard error as a function of deflated eigenvalues for the $J_4$ point-split operator at $\kappa = 0.1453$.}
\label{figs5}
\end{figure}

Subtraction methods were also employed to test their effectiveness on ensembles that include dynamical sea quarks. The calculations were performed on ten MILC configurations that use the HISQ gauge action, with $N_f = 2 + 1 + 1$ flavors, and a pion mass of 306.9(5) MeV on a lattice of size $16^3\times48$ \cite{MILC}. An analysis of the pion correlators for all ten configurations was performed to determine the value of the hopping parameter corresponding to the pion mass. Fig.~\ref{figs4} displays the fit to three different values of the hopping parameter, with the error bars calculated via jacknife analysis.

\begin{figure}
\centering
\includegraphics[trim={1cm 8cm 0 8cm},clip,width=1.10\textwidth]{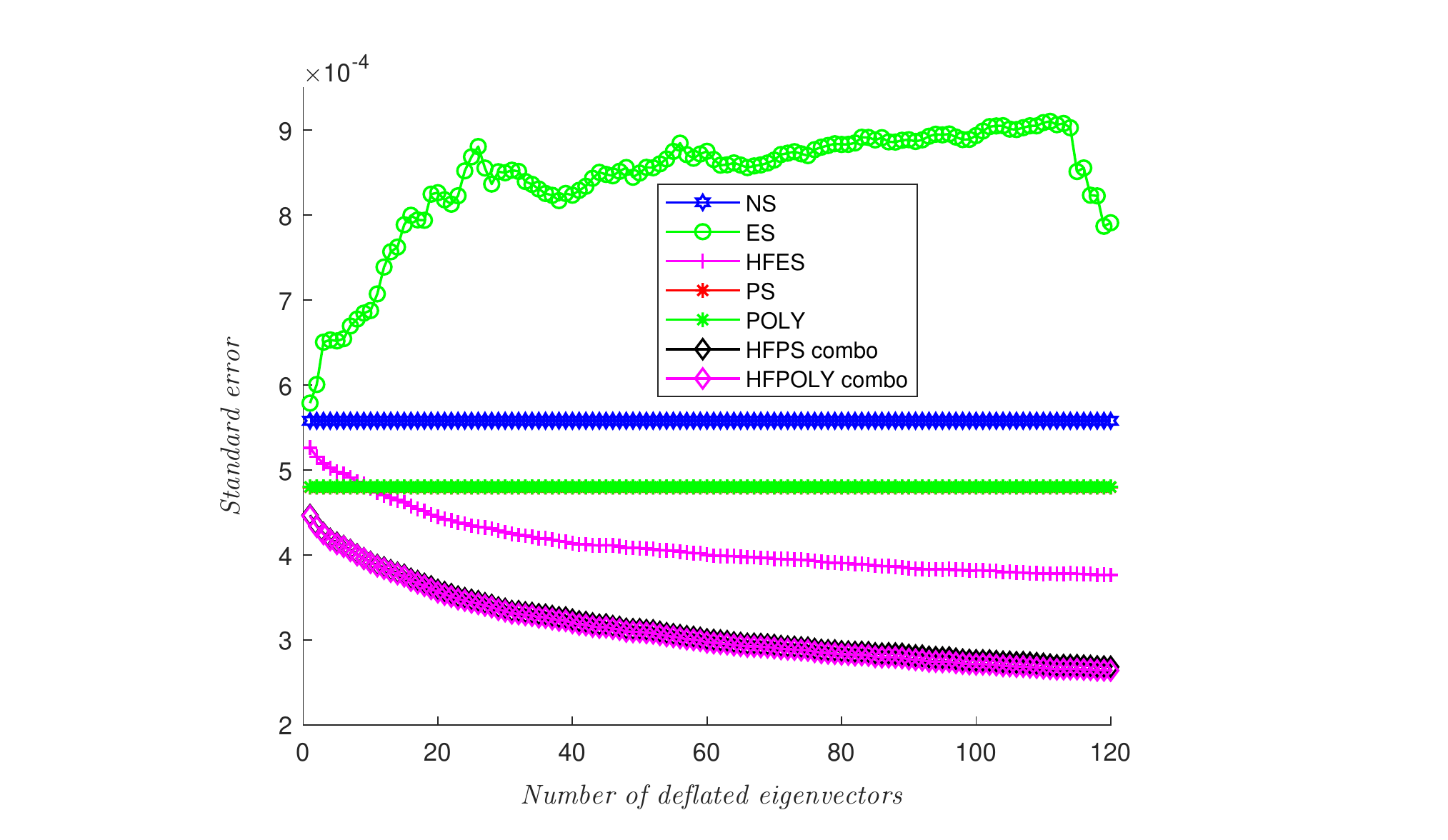}
\vspace{-.75cm}
\caption{Dynamical $16^3\times48$ lattice simulations for the standard error as a function of deflated eigenvalues for the $J_4$ local operator at $\kappa = 0.1453$.}
\label{figs6}
\includegraphics[trim={1cm 8cm 0 8cm},clip,width=1.10\textwidth]{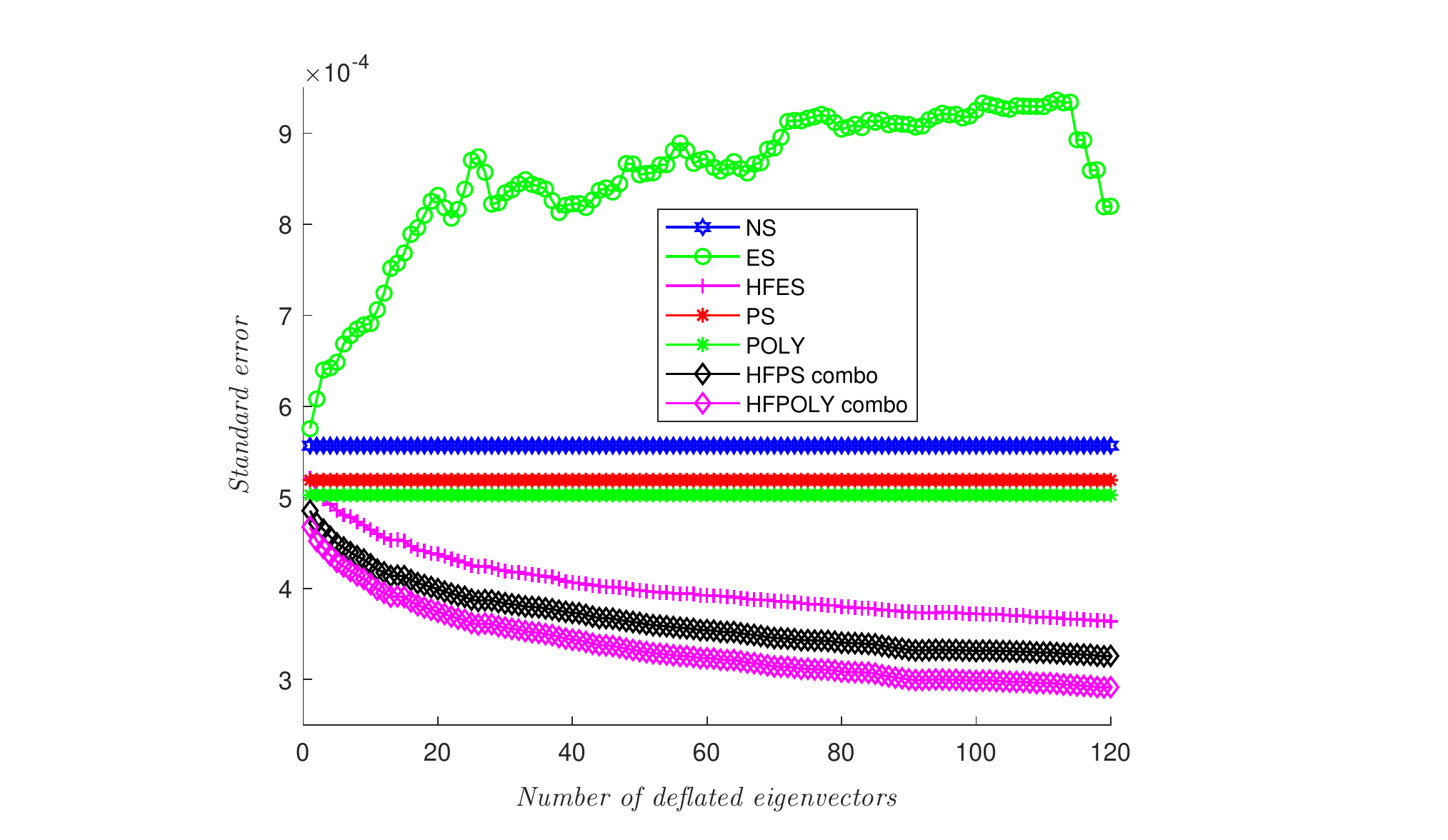}
\vspace{-.75cm}
\caption{Dynamical $16^3\times48$ lattice simulations for the standard error as a function of deflated eigenvalues for the scalar operator at $\kappa = 0.1453$.}
\label{figs7}
\end{figure}

Using the function obtained from the quadratic fit of the pion mass to the hopping parameter, the value of the hopping parameter corresponding to the mass of the pion for these ensembles was found to be $\kappa = 0.1453$, with $\kappa_{crit} = 0.1481$. The subtraction methods were then employed using this value of the hopping parameter, using eigenmodes of the Wilson matrix obtained from GMRES-DR, and eigenmodes of the hermitian matrix from MINRES-DR. The error bars were calculated for the point-split vector and local currents, as well as the scalar operator. Figs.~\ref{figs5}, \ref{figs6} and \ref{figs7} display the error bars as function of the number of deflated eigenmodes for each operator. Approximately 120 accurate eigenmodes were obtained for each configuration.

One thing evident in the dynamical case is that the ES results for all three operators display an extreme case of non-normality error value {\em increase}. This is in contrast with the milder quenched result increases and the actual decrease seen at $\kappa_{crit}$. Otherwise, the qualitative behavior of the point-split vector is similar to ones with lower values of $\kappa$ in the quenched approximation, although the deflation is more effective. The HFES result never goes below PS, for example in the earlier results. HFES is much more effective than PS and POLY for local vector and scalar operators. HFPOLY remains the most effective method for all operators while using dynamical lattices. Table \ref{table4} displays the relative efficiencies of each method in comparison to NS and PS. We also observe a small discontinuity at the first subtracted eigenvalue in these three figures. Our results demonstrate that our methods are effective in either the quenched or dynamical cases.

\begin{table}
\centering
\begin{tabular}{ c|c|c|c|c|c|c| } 
\cline{2-7}
\multicolumn{1}{c|}{}& \multicolumn{2}{c|}{Scalar}&\multicolumn{2}{c|}{Local $J_4$}&\multicolumn{2}{c|}{Point-Split $J_4$}\\
\hline
\multicolumn{1}{|c|}{Subtraction} & {\em{vs.}} NS  & {\em{vs.}} PS  &  {\em{vs.}} NS  & {\em{vs.}} PS &  {\em{vs.}} NS  &  {\em{vs.}} PS\\  
\hline
\multicolumn{1}{|c|}{POLY} & $22.8\%$ & $6.6\%$ & $35.0\%$ & $-0.1\%$ & $93.4\%$ & $5.2\%$   \\
\hline 
\multicolumn{1}{|c|}{HFES} & $134\%$ & $104\%$ & $120\%$ & $62.4\%$ & $60.0\%$ & $-13.2\%$  \\
\hline
\multicolumn{1}{|c|}{HFPS} & $192\%$ & $153\%$ & $332\%$ & $220\%$ & $417\%$ & $181\%$   \\
\hline
\multicolumn{1}{|c|}{HFPOLY} & $260\%$ & $217\%$ & $436\%$ & $230\%$ & $505\%$ & $229\%$  \\
\hline
\end{tabular}
 \caption{Comparison of relative efficiencies for operators at $\kappa=0.1453$ on the dynamical lattice of size $16^3\times48$.}
\label{table4}
\end{table}

Note that the MILC dynamical configurations, which have a pion mass of about 306.9 MeV, roughly corresponds to a quenched $\kappa\approx 0.1567$. This correspondence comes from a comparison of the physical pion mass in the MILC calculation\cite{MILC} and the Wilson fermions calculation in Ref.~\cite{Cabasino}. So, these dynamical lattices are intermediate between our quenched $24^3\times 32$ $\kappa=0.156$ lattices and our $\kappa_{crit}=0.157$ evaluations. This approximate correspondence is consistent with the small discontinuity seen at the first subtracted eigenvalue, intermediate between the almost continuous behavior at $\kappa=0.156$ seen in Figs.~\ref{figl64}, \ref{fignl64} and \ref{figl65} and the large discontinuities seen in Figs.~\ref{Tigs2}, \ref{Tigs1} and \ref{Tigs3} at $\kappa=0.157$. Thus, we expect even better results from deflation methods to occur for dynamical lattices at the physical point.

\section{Conclusions}\label{Sten}

Our polynomial and perturbative deflation combination methods produce very encouraging results. Our best results are for the deflation/polynomial method HFPOLY, which can be further extended through higher order implementation. We have seen that the efficiency of our deflation methods improve as we move on towards lower quark masses. From Table~\ref{table1} we see that an additional efficiency of over $\sim 1700\%$ and over $\sim 2000\%$ for local $J_4$ loop operators occurs at $\kappa_{crit}$ compared to standard PS and NS simulations, respectively. The use of the MINRES-DR algorithm has been crucial in this process. For dynamical lattices we observe deflation subtraction results consistent with the effectiveness seen in the quenched data.  

If we were running in production mode, other choices could be made for the computational setup. In particular, we use two special right hand side solves to get both the even-odd nonhermitian eigenmodes of the Wilson matrix, used for speeding up the other right hand sides using GMRES-Proj, as well as the eigenmodes of the full hermitian matrix, used for noise subtraction deflation. This could be simplified to a single special right hand side solve if we chose to speed up the hermitian calculation on the full Wilson matrix using MINRES-DR instead of GMRES-DR, and then re-use these for the noise subtraction. Such different production timings have not yet been explored. In addition, we have found a deflation saturation effect which shows that effective deflation can be done with only a few low eigenvalues, increasing the efficiency of a production mode setup. Our methodology employs deflation for both the linear equations solves and matrix noise subtraction. We use general matrix deflation algorithms, not specialized to lattice QCD, and straightforward noise deflation methods which are easy to implement.

Although we have used the Wilson action in most of our study, our methods may be generalized to many different lattice actions. Perturbative expansions may be applied to any generalized Wilson action such as clover or twisted types. In addition, deflation methods should be effective at low quark mass for any action. A rather strong low eigenvalue dominance has been established for operator variance in quenched LQCD near $\kappa_{crit}$. It is likely the same dominance will hold for dynamical lattices. As Monte Carlo lattice simulations move toward lighter dynamical quark masses, it will become more and more important to use deflation methods in the evaluation of quark loop effects.

There is much to explore beyond these results. As mentioned above, other lattice operators need to be partitioned in Dirac/color space for efficient simulations, and deflation on such partitions have not yet been considered. Physical point dynamical lattices need to be explored in detail. In addition, although we have combined POLY and deflation on the subtraction side, these methods have not yet been combined when solving the linear equations, nor have we yet considered POLY expansions of the hermitian system.

We would like to thank the Baylor University Research Committee and the Texas Advanced Computing Center for partial support. In addition we thank Abdou Abdel-Rehim, Victor Guerrero and Paul Lashomb for their contributions to this work. We would also like to thank Doug Toussaint, Carleton TeDar and Jim Hetrick who made MILC Collaboration gauge configurations available and assisted our analysis of them. We also thank Randy Lewis for his QQCD program. Calculations were done with HPC systems at Baylor University.
 

\newpage
\bibliography{referfile2}{}



%
%
%
%

\end{document}